\documentclass[twocolumn]{aastex631}

\usepackage{natbib}
\usepackage{wrapfig}
\usepackage{gensymb}
\usepackage{amsthm, amssymb, amsfonts} 
\usepackage{physics}   
\usepackage{wasysym}
\usepackage{CJK}

\begin{document}
\begin{CJK*}{UTF8}{gbsn}

\title{Surveying Ultra-hot Jupiters using Phase Curves with \textit{Twinkle}}

\author[0000-0002-0786-7307]{Kaz Gary}
\affiliation{The Ohio State University \\
140 West 18th Avenue \\
Columbus, OH 43210, USA}

\author[0000-0002-4361-8885]{Ji Wang (王吉)}
\affiliation{The Ohio State University \\
140 West 18th Avenue \\
Columbus, OH 43210, USA}

\author[0000-0002-8823-8237]{Anusha Pai Asnodkar}
\affiliation{The Ohio State University \\
140 West 18th Avenue \\
Columbus, OH 43210, USA}
\affiliation{California Institute of Technology \\
1200 East California Boulevard \\
Pasadena, CA 91125, USA}

\author[0000-0001-9665-8429]{Ian Wong}
\affiliation{Space Telescope Science Institute \\
3700 San Martin Drive \\
Baltimore, MD 21218, USA}

\begin{abstract}

Due to their high equilibrium temperatures ($T_{eq}$ $>$ 2000 K), ultra-hot Jupiters (UHJs) are the best characterized exoplanets to date. However, many questions about their formation, evolution, and atmospheres remain unanswered. Phase curve observations can reveal answers to these questions by constraining multiple atmospheric properties including circulation, albedo, and chemistry. To this end, we simulate and forecast a survey of UHJ atmospheres via phase curve observations with the upcoming \textit{Twinkle} mission. \textit{Twinkle} is a spectroscopic satellite covering 0.5--4.5 $\micron$ with a spectral resolving power of R $\sim$ 50--70. Using a physically motivated model, we simulate white-light photometric phase curve observations for 14 UHJs in \textit{Twinkle's} field of regard. We project that \textit{Twinkle} will be able to detect all phase curve signals in our survey. Additionally, we simulate spectroscopic phase curves for the UHJ, WASP-189~b. From our simulated spectroscopic phase curves, we generate mock phase-resolved emission spectra. Previously detected UHJ molecules (e.g. H$_2$O, CO and CO$_2$) produce notable features in the resulting spectra, allowing for detailed atmospheric characterization to study the 3D structure of UHJ atmospheric chemistry and dynamics. For planets with hotspot phase offsets, \textit{Twinkle} will be capable of detecting them both in the optical and infrared wavelength ranges. This future survey would represent the first UHJ phase curve survey with simultaneous coverage in optical and infrared wavelengths and will provide new constraints and reveal intriguing trends in these extreme environments.

\end{abstract}


\section{Introduction} \label{sec:intro}

Since the first detection of an atmosphere around an extrasolar planet \citep{Charbonneau2002}, numerous studies have been dedicated to understanding exoplanetary atmospheric dynamics and composition (\citealt{atmospherereview} and references therein). Among future missions (e.g. \citealt{Arielexoplanets}, \citealt{pandoramission}, \citealt{Plato}), the \textit{Twinkle} mission will be one of the first space-based missions dedicated to spectroscopic characterization of exoplanet atmospheres \citep{twinkle_original_paper}. \textit{Twinkle} is a 0.45 m space-based telescope with an estimated launch date of October 2027 that will provide low resolution spectroscopy within the visible and infrared wavelength ranges (0.5--4.5 \micron) across two channels. With close to 1000 exoplanets within \textit{Twinkle}'s field of regard in the ecliptic plane \citep{caprice_twinkle}, systematic population surveys of exoplanet atmospheres will be possible.

One of the most comprehensive ways to characterize an exoplanet's atmosphere is through orbital phase curves. An orbital phase curve, or phase curve, is the extracted planetary signal from observing a star-planet system across the planet's entire orbit. If the planet is transiting, phase curves can capture the primary transit as well as the secondary eclipse, when the planet is occulted by the host star.

For a tidally-locked, transiting planet, the changing viewing phase during its orbital period creates a minimum in the observed planetary flux around the primary transit and a maximum around the secondary eclipse. This maximum can be offset from the center of the secondary eclipse and is typically the substellar point. A ``phase offset" in the maximum of a phase curve would imply that the brightest region on the planet has been offset either eastward or westward of the substellar point. Depending on the wavelength at which the phase curve is observed, both of these parameters offer different insights into the planet's atmosphere. Optical wavelengths probe the reflected light from the planet, providing a 2D longitudinal map of the albedo. In infrared wavelengths, thermal emission from the planet can probe the atmospheric chemical composition and temperature ~\citep[see][for a full review on the planetary phase curve signal]{crossfield2018}.

Ultra-hot Jupiters (UHJs) are the most amenable targets for phase curve measurements. Their large radii and high temperatures result in high-amplitude phase curves. Their short periods make it convenient for repeated observations to improve the signal-to-noise ratio (SNR). Previous observations of UHJ phase curves have been taken in the optical and near-infrared wavelengths with the Transiting Exoplanet Survey Satellite (TESS) and the Hubble Space Telescope (HST) \citep[e.g.,][]{WASP-167b2, Wong2020kelt9, Wong2021} as well as at thermal infrared wavelengths with the \textit{Spitzer} Space Telescope \citep[e.g.,][and references therein]{Dang2024, spitzerphasecurvestudy} and JWST \citep[e.g.,][]{TRAPPISTJWST, WASP121phasecurve}. No telescope to date has simultaneously observed UHJs at optical and thermal infrared wavelengths. With \textit{Twinkle}, a homogeneous, multi-wavelength data set of UHJ phase curves can be obtained for the first time. This data set would allow for direct comparison between observations and would comprehensively probe UHJ atmospheric dynamics, structure, and chemical composition.

Additionally, the observatory will be complementary to JWST. While JWST's data quality is unprecedented, there is limited time available for phase curve population surveys. \textit{Twinkle} can fill in this gap by conducting initial atmospheric observations, including a dedicated population survey, and offer evidence to motivate higher-resolution follow-up with JWST on the most interesting targets. JWST and \textit{Twinkle} combined will maximize the scientific outputs of both missions and contribute greatly to the exoplanet atmosphere community.

In this work, we explore the capability and science output of a dedicated UHJ phase curve survey as part of \textit{Twinkle's} broader exoplanet survey. Our sample of fourteen targets covers confirmed gas giant planets (planetary masses $<$ 15 Jupiter masses) with equilibrium temperatures greater than 2000 K. The survey represents the first systematic survey of UHJ phase curves that simultaneously covers optical and thermal wavelengths.

The paper is structured as follows. We first present our target selection criteria in Section \ref{sec:target selection}. We then describe the physically-motivated model used to simulate phase curve data in Section \ref{sec:methods}. We present mock observations and the projected yield of a UHJ survey with \textit{Twinkle} in Section \ref{sec:results} and discuss what possible science cases we could explore with this data in Section \ref{sec:discussion}. Finally, we present our conclusions and recommendations for a dedicated UHJ phase curve survey with \textit{Twinkle} in Section \ref{sec:conclusions}.

\section{Target Selection} \label{sec:target selection}

We first selected targets for our survey based on equilibrium temperature, planet mass, and a metric for the predicted \textit{Twinkle} signal-to-noise ratio (SNR) of each planet's phase curve. To do this, we queried the NASA Exoplanet Archive for confirmed planets with equilibrium temperatures greater than 2000 K and masses less than 15 $M_J$. The mass cut-off was used to exclude high gravity objects from our sample. We used the default parameter set in our simulations to ensure consistency.

This query returned $\sim$60 planets. We calculated a preliminary ``binned" SNR by collapsing a full orbit light curve into a single point as a metric of the predicted phase curve precision from \textit{Twinkle} for all planets using the following procedure.

\subsection{Signal Calculation} \label{susbsubsec:noise}

For our signal, we used a flux ratio between the planet and the star to represent the phase curve signal,

\begin{linenomath}
\begin{equation}
\label{eq:SNR_S}
    S = \frac{T_{eq, p}^4 \times R_{p}^2}{T_{*}^4 \times R_{*}^2}
\end{equation}
\end{linenomath}

where $T_{eq, p}$ is the equilibrium temperature of the planet, $R_{p}$ is the radius of the planet, $T_{*}$ is the temperature of the host star and $R_{*}$ is the radius of the host star. These values were the default parameters reported for each planet in the NASA Exoplanet Archive. Default parameters are determined by multiple factors including the completeness of the parameter set from a single source and the precision of reported parameters. For more information on the default parameter criteria, see \citealt{Christansen25}.

\subsection{Photon Noise Calculation} \label{susbsubsec:noise}

Noise in our observations will be dominated by photon noise from the host star. We first calculated photon noise from the host star by assuming the star would emit as a blackbody and scaled this radiation by exposure time and \textit{Twinkle}'s aperture (0.45 m). We used half of each planet's orbital period as our exposure time. This is to simulate the minimum $50 \%$ overhead of \textit{Twinkle}'s observations over each wavelength bin in \textit{Twinkle's} wavelength range. This represents a half-orbit long stare of the system, which is long enough to capture the planet's phase curve form maximum to minimum. At each of these wavelengths, we used the following equation to calculate the star's emission in units of ergs,

\begin{equation} \label{eq:staremission}
    f = \sum_{i = 0}^{M} B_{\lambda, T} \times \Omega^2 \times (\pi \times \left(\frac{D}{2}\right)^2) \times t_{exp} 
\end{equation}

where $B_{\lambda, T}$ is the blackbody radiation with respect to wavelength and stellar temperature, $D$ is \textit{Twinkle}'s primary mirror diameter and $t_{exp}$ is the exposure time. $\Omega$ is the solid angle and is equivalent to $(\arctan(\frac{R_P}{d_{sys}}))^2$ where $R_P$ is the planet radius and $d_{sys}$ is the distance to the system. For a visual representation, see Figure \ref{fig:modeldiagram}.

We calculated $f$ in every wavelength channel within \textit{Twinkle}'s wavelength range and then summed these values to get the total white light radiation detected by \textit{Twinkle} from the star in ergs. We then converted this value into the total number of individual photons ($\gamma$) by,

\begin{equation}
\label{eq:N_1}
    \gamma = \frac{f}{\frac{hc}{\lambda}}
\end{equation}

and calculated total photon noise using the following equation,

\begin{equation}
\label{eq:Twinkle_noise_calc}
    N = \frac{1}{\sqrt{\gamma}}
\end{equation}

\textit{Twinkle's} noise floor is about 20 parts-per-million (ppm) \citep{main_twinkle_paper}. If this number was below the noise floor, then we used 20 ppm as our noise ($N$) in Equation \ref{eq:twinkle_SNR}.

To calculate our estimated \textit{Twinkle} SNR of the white-light phase curve we used the following equation,

\begin{equation}
\label{eq:twinkle_SNR}
    SNR_{Twinkle} = \frac{S}{N}
\end{equation}

where S is defined in Equation \ref{eq:SNR_S} and N is the larger value between Equation \ref{eq:Twinkle_noise_calc} or 20 ppm.

Once we calculated the \textit{Twinkle} SNR for each planet, we aimed to identify $\sim$20 gas giant planets with an SNR $\geq$ 2 that were in \textit{Twinkle's} field of regard which focuses on the ecliptic plane. We chose twenty as our ideal sample size, because previous work on ultra-hot Jupiter phase curve surveys were able to draw trends from a population of this size (e.g. \citealt{Wong2020b}, \citealt{Wong2021}, \citealt{Bell2021}, \citealt{Dang2024}). We use 2 as our SNR cut-off to limit our sample size to be within 20 for the practical limitation of available Twinkle time for the survey. We emphasize that SNR in this section is purely an estimation of the observation metric rather than the actual SNR calculation which will be detailed in Section \ref{sec:methods}. We utilized the \textit{Twinkle} orbital planning tool developed by the \textit{Twinkle} team to confirm if planets would be in its field of regard.

In the end, 14 planets were in \textit{Twinkle's} field of regard, were classified as gas giant planets and had calculated SNR's high enough to have measurable phase curve signals. For the 14 targets in our survey, we estimated the total survey time to be $\sim$50 days. This is double the amount of time of all the planets' orbital periods ($\sim$25 days) to provide a conservative estimate of overheads since the final configuration of \textit{Twinkle}'s orbit has yet to be determined. The first year ($\sim$5600 hours) of \textit{Twinkle}'s mission will be dedicated to exoplanet surveys, with split time in subsequent years of its primary 3 year mission. Our phase curve survey will amount to 20 percent of this dedicated time in the first year.

\section{Methods} \label{sec:methods}

\subsection{Phase Curve Model} \label{subsec:phase curve model}

Our full phase curve model includes effects from the planet as well as all stellar contributions in the phase curve signal. These plenatary effects include the thermal emission from the planet itself and reflected stellar light from the planet's atmosphere. We also incorporate secondary effects from the star such as ellipsoidal variation and Doppler boosting. Modeling of the secondary eclipse and primary transit are not included in this paper because we are primarily interested in the day-night temperature contrast and the phase offset, which are revealed by the phase curve modulation.

\begin{figure*}[th!]
\centering
\includegraphics[width=0.5\textwidth]{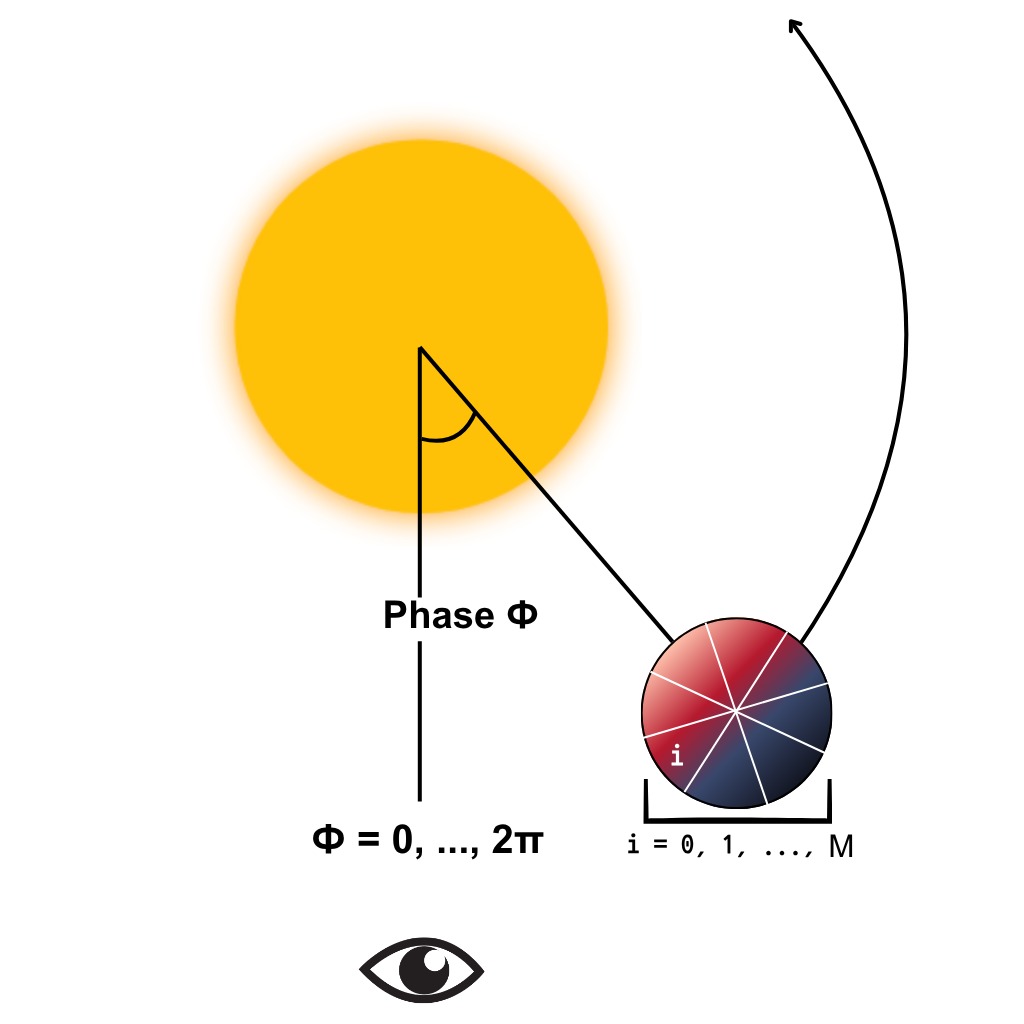}
\caption{Diagram of our full phase curve model described by Eq. \ref{eq:fullphasecurve}. $\phi$ represents the angle between the observer's line of sight (at phase 0) and the phase at which the planet is at. The planet is split up into M longitudinal slices and each slice is represented by a number, i. 
\label{fig:modeldiagram}}
\end{figure*}

Our model divides the planet into equal longitudinal slices and we assume a tidally-locked planet with a permanent day- and night-side. For our simulations, we divided the planet into sixteen slices. At a given phase, the thermal emission and reflected light components are calculated for each observable slice. All observable slices are then summed together to produce the planetary signal at each phase. The calculations for these components are described in Section \ref{susbsubsec:thermal} and Section \ref{susbsubsec:reflection}. The secondary effects of the host star are calculated and added to this sum at each phase. The calculations for this component are described in Section \ref{susbsubsec:secondary harmonics}. A diagram of our model is shown in Figure \ref{fig:modeldiagram}. 

\subsubsection{Thermal Emission} \label{susbsubsec:thermal}

In thermal infrared wavelengths, the minimum of the phase curve corresponds to the hemispherically averaged brightness of the night-side while the maximum corresponds to the hemispherically averaged brightness of the day-side. Under the assumption of blackbody radiation, the brightness of each side can be converted to a temperature. Thus, the amplitude of a thermal phase curve signal corresponds to the day-night temperature contrast of the planet. Additionally, the maximum of the thermal phase curve, whether offset from the substellar point or not, corresponds to the hottest region of the planet.

To model the thermal emission at a given phase, we use the following equation at each slice to calculate its individual thermal contribution,

\begin{equation} \label{eq:thermal at each phase}
    f_{thermal,\, i} = 1.0 - f_{drop} \times \left( \frac{i}{M/2}\right)^{\alpha}
\end{equation}

where $f_{drop}$ is the flux drop from the day to night side, $i$ is the slice's index at the current phase being simulated, M is the total number of planet slices, and $\alpha$ is the interpolation power that maps the planet's thermal distribution between slices. The slice closest to the star (i.e. the substellar point) has an index of zero. The thermal emission gradient is then calculated from the day-side to the night-side based upon the interpolation power and then mirrored on the other side to simulate a symmetric heat distribution. The ratio, $\left( \frac{i}{M/2}\right)$ is used to ensure that the dayside flux ($i$ = 0) is normalized to 1 as an arbitrary value and the nightside flux ($i$ = $M/2$) is normalized to the (1 - $f_{drop}$) parameter. 

We then sum all of the observable slices at a given phase to get the total thermal contribution from the planet at each phase,

\begin{equation} \label{eq:thermal}
    f_{thermal} = \sum_{i \in \mathrm{\{visible\ slices\}}} f_{thermal,\, i}
\end{equation}

To account for the planet's thermal phase offset, we add the offset (in radians) to the phase angle to ensure a maximum in the phase curve occurs at this offset.

\subsubsection{Reflected Light} \label{susbsubsec:reflection}

To model the host star's reflected light from the planet, we assume the planet is a Lambertian reflector. The Lambertian reflectance formula is,

\begin{equation} \label{eq:reflection}
    f_{reflection} = \sin{\phi} + (\pi - \phi) \times \cos{\phi},
\end{equation}
where $\phi$ is the phase angle as defined from 0 to $2\pi$ where phase 0 is mid-transit and $\pi$ is the middle of the secondary eclipse \citep{lambertianreflectance}. The $f_{reflection}$ at each phase is then multiplied by the amplitude, $\epsilon$, of the scaled geometric albedo, $A_g$. This is defined as,

\begin{equation}
\label{eq:geometric albedo}
    \epsilon = A_g \times \frac{R_p^2}{a^2}
\end{equation}

where $R_p$ is the radius of the planet and $a$ is the semi-major axis. At each phase angle, the thermal emission and reflected starlight of the observable planet slices are summed.

\subsubsection{Secondary Effects} \label{susbsubsec:secondary harmonics}

Beyond the planetary contribution to the phase curve signal, the host star can also contribute to the phase signal if the planet mass is sufficiently high. One of these effects is Doppler boosting, which occurs when the observed flux is periodically amplified due to the radial velocity of the host star. The Doppler boosting signal has a maximum at quadrature when the star is moving towards the observer and a minimum at quadrature when the star is moving away from the observer. The resulting photometric variation signal can be best described by a sinusoidal variation with a period matching the planet's orbital period but offset in phase. The other effect is ellipsoidal distortion, which occurs when the planet's gravity distorts the shape of the host star. The ellipsoidal distortion signal reaches a maximum at the two quadratures when the observer sees the elongated side of the distorted star and reaches a minimum at the conjunctions when the observer sees the shorter side of the distorted star. The resulting photometric variation signal can be best described by a sinusoidal wave with half of the orbital period ~\citep[see][for a full review on these effects]{Shporer2017}.

To model these effects, we use the following equation adapted from Eq. 2 in \citep{Wong2020b},

\begin{equation} \label{eq:harmonics}
    f_{harmonics} = A_1 \sin(\phi) - A_2 \cos(2\phi)
\end{equation}

where $A_1$ is the Doppler boosting amplitude, $A_2$ is the ellipsoidal distortion signal amplitude and $\phi$ is the phase angle. We used Equation 4 and Equation 7 in \cite{Shporer2017} to calculate the theoretical values of $A_1$ and $A_2$, respectively. Values for each of the planets are reported in Table \ref{tab:harmonics}. Any signal below \textit{Twinkle's} noise floor will not be detectable. All calculated Doppler boosting signals were below \textit{Twinkle}'s noise floor. If the median of the planet's possible ellipsoidal distortion amplitude exceeded this threshold, we listed the planet's name in bold in Table \ref{tab:harmonics} and included these effects in our modeling of the planet. All other planets in our survey were modeled using only the thermal and reflection components of our full phase curve model.

\subsubsection{Full Phase Curve Model} \label{susbsubsec:fullphasecurve}

Our final full phase curve model then becomes,

\begin{equation} \label{eq:fullphasecurve}
    f_{pc}(\phi) = (f_{thermal} / f_{star})\ + (f_{reflection} \times \epsilon) + f_{harmonics}
\end{equation}

where $f_{star}$ is defined as the the ratio of the star's emission to the planet's signal at each phase as,

\begin{equation} \label{eq:thermal2}
    f_{star} = (B_{sp} \times R_{sp}^2) \times (M / 2)
\end{equation}

where $B_{sp}$ is the blackbody radiation ratio between the star and planet at the observed wavelength using the star's effective temperature and the temperature of the planet's hottest slice, $R_{sp}$ is the radius ratio between the star and planet, and $M$ is the total number of slices of the planet.

In the thermal component of our model (\ref{eq:thermal}) we sum over only the observable slices of the planet. Therefore, since the thermal emission of each slice is normalized to 1 (\ref{eq:thermal at each phase}) and there in total $M/2$ observable slices at a given phase, the thermal emission of the planet is normalized to $M/2$. To ensure consistency with the stellar component, we multiply by $M/2$ such that the star-planet thermal emission flux ratio is scaled correctly. The total thermal contribution is therefore the sum of the planet's visible slices at each phase (Equation $\ref{eq:thermal}$) divided by the total thermal emission (Equation $\ref{eq:thermal2}$) from the star. 

\begin{deluxetable*}{lccc}
\tabletypesize{\scriptsize}
\tablewidth{0pt} 
\tablecaption{Theoretical Harmonics Values for All Planets \label{tab:harmonics}}
\tablehead{
\colhead{Planet} & \colhead{Ellipsoidal Distortion amplitude}& \colhead{Doppler Boosting amplitude} & \\
\colhead{}& \colhead{(ppm)} &  \colhead{(ppm)}
} 
\startdata
\textbf{WASP-33b} & 36 - 55 & 4 \\ 
\textbf{TOI-2109b} & 311 - 480 & 11 \\ 
WASP-76b & 11 - 17 & 1.6 \\ 
WASP-189b & 12 - 18 & 2.4 \\ 
\textbf{WASP-167b} & 76 - 118 & 11 \\ 
\textbf{HAT-P-23b} & 27 - 43 & 5.0 \\ 
\textbf{HAT-P-70b} & 25 - 38 & 8.5 \\ 
Wendelstein-1b & 1.0 - 1.6 & 1.5 \\ 
\textbf{WASP-103b} & 55 - 86 & 3.6 \\ 
\textbf{HATS-24b} & 28 - 43 & 4.9 \\ 
WASP-178b & 3 - 6 & 1.9 \\ 
Wendelstein-2b & 2.6 - 4.0 & 2.0 \\ 
KELT-17b & 3 - 5 & 1.6 \\ 
HD 202772Ab & 8 - 13 & 1.2 \\ 
\enddata
\tablecomments{List of all estimated secondary harmonics effects for all planets. The planets marked in bold have these effects modeled in their phase curves. Note that the ellipsoidal distortion amplitude is a range, given the reasonable gravity darkening parameters (0.3 - 1).}
\end{deluxetable*}

\subsection{Data Simulation} \label{subsec:data simulation}

\subsubsection{Photometric Phase Curve Simulation} \label{subsubsec:photometric phase curve sim}

We simulated photometric phase curve data for all planets in our survey. The list of parameters used for each planet can be found in Tables \ref{tab:planetparameters1} and \ref{tab:planetparameters2} with the preliminary SNR value used in target selection (\ref{eq:twinkle_SNR}) and our retrieved SNR after simulating the data. The retrieved SNRs were calculated using the 50th percentile of the posterior distribution for the phase curve amplitudes and dividing by the simulated photon noise. For planets listed in bold in Table \ref{tab:harmonics}, the median of the possible distribution of ellipsoidal amplitudes was used as the planet's ellipsoidal distortion amplitude in data simulation.

\defcitealias{HAT-P-23b}{[1]}
\defcitealias{HAT-P-70b1}{[2]}
\defcitealias{HAT-P-70b2}{[3]}
\defcitealias{HATS-18b}{[4]}
\defcitealias{HATS-24b}{[5]}
\defcitealias{HD202772Ab}{[6]}
\defcitealias{K2-260b}{[7]}
\defcitealias{Qatar-7b}{[8]]}
\defcitealias{KELT-17b1}{[9]}
\defcitealias{KELT-17b2}{[10]}
\defcitealias{TOI-2109b}{[11]}
\defcitealias{WASP-33b1}{[12]}
\defcitealias{WASP-33b2}{[13]}
\defcitealias{WASP-76b1}{[14]}
\defcitealias{WASP-76b2}{[15]}
\defcitealias{WASP-76b3}{[16]}
\defcitealias{WASP-103b1}{[17]}
\defcitealias{WASP-103b2}{[18]}
\defcitealias{WASP-103b3}{[19]}
\defcitealias{WASP-114b}{[20]}
\defcitealias{WASP-167b1}{[21]}
\defcitealias{WASP-167b2}{[22]}
\defcitealias{WASP-178b}{[23]}
\defcitealias{WASP-189b1}{[24]}
\defcitealias{WASP-189b2}{[25]}
\defcitealias{WASP189bTP}{[26]}
\defcitealias{Wendelsteinplanets}{[27]}

\begin{deluxetable*}{lcccccc}
\tabletypesize{\scriptsize}
\tablewidth{0pt} 
\renewcommand{\arraystretch}{1.7}
\tablecaption{Planet Parameters used in Data Simulation \label{tab:planetparameters1}}
\tablehead{
\colhead{Planet} & \colhead{} & \colhead{Albedo amplitude, $\epsilon$} & \colhead{Flux drop, $f_{drop}$} & \colhead{Thermal Offset, $\delta$} & \colhead{Interpolation Power, $\alpha$} & \colhead{Blackbody ratio, $B_{sp}$} \\ \colhead{ } & \colhead{} & \colhead{(ppm)}& \colhead{ } & \colhead{(radians)} & \colhead{ } & \colhead{ }
} 
\startdata
WASP-33b & \textit{Input:} & 0 & 0.93 \citetalias{WASP-33b1} & 0.5 \citetalias{WASP-33b1} & 1.0 & 9.7 \citetalias{WASP-33b1}\\
{       }& \textit{Retrieved:} & $54.3^{+20}_{-21}$ & $0.8^{+0.008}_{-0.007}$ & $0.5^{+0.02}_{-0.01}$ & $0.98^{+0.03}_{-0.02}$ & $9.7^{+0.2}_{-0.2}$ \\
\hline
TOI-2109b & \textit{Input:} & 0 & 0.77 \citetalias{TOI-2109b} & 0 & 1.0 & 4.33 \citetalias{TOI-2109b}\\
{       }& \textit{Retrieved:} & $283^{+64.7}_{-148}$ & $0.77^{+0.11}_{-0.04}$ & $0.007^{+0.01}_{-0.01}$ & $0.72^{+0.16}_{-0.16}$ & $4.55^{+0.64}_{-0.58}$\\
\hline
WASP-76b & \textit{Input:} & 72.8 \citetalias{WASP-76b2} & 0.87 \citetalias{WASP-76b1} & 0 \citetalias{WASP-76b3} & 1.0 & 7.4 \citetalias{WASP-76b1}\\
{       }& \textit{Retrieved:} & $86.6^{+45.5}_{-38.7}$ & $1.2^{+0.01}_{-0.01}$ & $-0.19^{+0.01}_{-0.01}$ & $1.02^{+0.07}_{-0.07}$ & $9.7^{+0.2}_{-0.2}$ \\
\hline
WASP-189b & \textit{Input:} & 61.0 \citetalias{WASP-189b2} & 1.0 & 0 & 1.0 & 9.5 \citetalias{WASP-189b1}\\
{       }& \textit{Retrieved:} & $155^{+9.32}_{-21.7}$ & $0.74^{+0.02}_{-0.02}$ & $0.02^{+0.009}_{-0.009}$ & $1.74^{+0.16}_{-0.26}$ & $8.9^{+1.1}_{-0.97}$ \\
\hline
WASP-167b & \textit{Input:} & 153.1 \citetalias{WASP-167b2} & 0.85 \citetalias{WASP-167b2} & -0.3 \citetalias{WASP-167b2} & 1.0 & 24.8 \citetalias{WASP-167b2} \\
{       }& \textit{Retrieved:} & $242^{+23}_{-97}$ & $0.87^{+0.1}_{-0.02}$ & $-0.39^{+0.1}_{-0.01}$ & $0.72^{+0.16}_{-0.16}$ & $12.9^{+0.6}_{-1.02}$\\
\hline
HAT-P-23b & \textit{Input:} & 0 & 1.0 & 0.48 \citetalias{HAT-P-23b} & 1.0 & 20.5 \citetalias{HAT-P-23b} \\
{       }& \textit{Retrieved:} & $123^{+51}_{-76}$ & $0.75^{+0.07}_{-0.04}$ & $0.7^{+0.12}_{-0.18}$ & $1.3^{+0.23}_{-0.27}$ & $20.3^{+0.83}_{-1.0}$\\
\hline
HAT-P-70b & \textit{Input:} & 0 & 1.0 & 0 & 1.0 & 34.4 \citetalias{HAT-P-70b2} \\
{       }& \textit{Retrieved:} & $18.7^{+16}_{-11.5}$ & $1.02^{+0.09}_{-0.06}$ & $0.01^{+0.02}_{-0.03}$ & $1.34^{+0.4}_{-0.3}$ & $34.8^{+1.75}_{-1.33}$\\
\hline
Wendelstein-1b & \textit{Input:} & 0 & 1.0 & 0 & 1.0 & 23.4 \citetalias{Wendelsteinplanets}\\
{       }& \textit{Retrieved:} & $205^{+52}_{-78}$ & $0.72^{+0.05}_{-0.04}$ & $0.03^{+0.03}_{-0.03}$ & $1.48^{+0.3}_{-0.3}$ & $23.4^{+1.1}_{-1.1}$\\
\hline
WASP-103b & \textit{Input:} & 0 \citetalias{WASP-103b3} & 0.84 \citetalias{WASP-103b2} & 0 \citetalias{WASP-103b2} & 1.0 & 7.3 \citetalias{WASP-103b2}\\
{       }& \textit{Retrieved:} & $184^{+187}_{-143}$ & $0.9^{+0.11}_{-0.13}$ & $-0.01^{+0.04}_{-0.06}$ & $0.94^{+0.62}_{-0.30}$ & $7.43^{+0.51}_{-0.48}$ \\
\hline
HATS-24b & \textit{Input:} & 0 & 1.0 & 0 & 1.0 & 29.7 \citetalias{HATS-24b} \\
{       }& \textit{Retrieved:} & $60.8^{+49.7}_{-41.1}$ & $0.88^{+0.12}_{-0.08}$ & $0.05^{+0.05}_{-0.05}$ & $0.97^{+0.62}_{-0.31}$ & $29.8^{+0.85}_{-1.01}$\\
\hline
WASP-178b & \textit{Input:} & 57.0 \citetalias{WASP-178b} & 0.75 \citetalias{WASP-178b} & 0 & 1.0 & 41.2 \citetalias{WASP-178b} \\
{       }& \textit{Retrieved:} & $22.1^{+19.2}_{-14.6}$ & $1.17^{+0.05}_{-0.10}$ & $-0.01^{+0.01}_{-0.01}$ & $1.75^{+0.16}_{-0.41}$ & $40.1^{+4.98}_{-3.66}$ \\
\hline
{Wendelstein-2b} & \textit{Input:} & 0 & 1.0 & 0 & 1.0 & 20.8 \citetalias{Wendelsteinplanets} \\
{       }& \textit{Retrieved:} & $158.8^{+129}_{-92.2}$ & $0.92^{+0.13}_{-0.10}$ & $0.01^{+0.03}_{-0.05}$ & $1.10^{+0.61}_{-0.37}$ & $20.7^{+0.99}_{-0.98}$ \\
\hline
{KELT-17b} & \textit{Input:} & 0 & 1.0 & 0 & 1.0 & 59.0 \citetalias{KELT-17b1} \\
{       }& \textit{Retrieved:} & $4.71^{+4.76}_{-3.73}$ & $0.94^{+0.02}_{-0.02}$ & $-0.01^{+0.01}_{-0.02}$ & $0.74^{+0.15}_{-0.11}$ & $58.9^{+0.62}_{-0.47}$ \\
\hline
{HD 202882Ab} & \textit{Input:} & 0 & 1.0 & 0 & 1.0 &  32.0 \citetalias{HD202772Ab} \\
{       }& \textit{Retrieved:} & $4.64^{+5.26}_{-2.93}$ & $0.91^{+0.05}_{-0.03}$ & $-0.06^{+0.03}_{-0.04}$ & $0.88^{+0.26}_{-0.22}$ & $32.6^{+1.17}_{-1.16}$ \\
\enddata
\tablerefs{1, \cite{HAT-P-23b}; 2, \cite{HAT-P-70b1}; 3, \cite{HAT-P-70b2}; 4, \cite{HATS-18b}; 5, \cite{HATS-24b}; 6, \cite{HD202772Ab}; 7, \cite{K2-260b}; 8, \cite{Qatar-7b}; 9, \cite{KELT-17b1}; 10, \cite{KELT-17b2}; 11, \cite{TOI-2109b}; 12, \cite{WASP-33b1}; 13, \cite{WASP-33b2}; 14, \cite{WASP-76b1}; 15, \cite{WASP-76b2}; 16, \cite{WASP-76b3}; 17, \cite{WASP-103b1}; 18, \cite{WASP-103b2}; 19, \cite{WASP-103b3}; 20, \cite{WASP-114b}; 21, \cite{WASP-167b1}; 22, \cite{WASP-167b2}; 23, \cite{WASP-178b}; 24, \cite{WASP-189b1}; 25, \cite{WASP-189b2}; 26, \cite{WASP189bTP}; 27, \cite{Wendelsteinplanets}}
\end{deluxetable*}

\begin{deluxetable*}{lcccccc}
\tabletypesize{\scriptsize}
\tablewidth{0pt} 
\renewcommand{\arraystretch}{1.7}
\tablecaption{Planet Parameters used in Data Simulation (cont.) \label{tab:planetparameters2}}
\tablehead{
\colhead{Planet} & \colhead{} & \colhead{Radius ratio, $R_{sp}$} & \colhead{Doppler boosting amplitude, $A_1$} & \colhead{Ellipsoidal distortion amplitude, $A_2$} & \colhead{Preliminary SNR} & \colhead{Retrieved SNR} \\ \colhead{(cont.)} & \colhead{ } & \colhead{ } & \colhead{(ppm)} & \colhead{(ppm)} & \colhead{ } & \colhead{ }
} 
\startdata
WASP-33b & \textit{Input:} & 8.6 \citetalias{WASP-33b1} & 4 & 33 & 12 & \nodata\\
{       } & \textit{Retrieved:} & $7.1^{+0.1}_{-0.1}$ & $4.1^{+1.0}_{-0.87}$ & $38^{+1.2}_{-1.5}$ & \nodata & 54\\
\hline
TOI-2109b & \textit{Input:} & 12.3 \citetalias{TOI-2109b} & 11 & 419 & 32 & \nodata\\
{       } & \textit{Retrieved:} & $12.9^{+1.22}_{-0.65}$ & $11.5^{+1.05}_{-0.62}$ & $440^{+4.2}_{-11.5}$ & \nodata & 44\\
\hline
WASP-76b & \textit{Input:} & 7.6 \citetalias{WASP-76b1} & 0 & 0 & 9 & \nodata\\
{       } & \textit{Retrieved:} & $7.1^{+0.1}_{-0.1}$ & \nodata & \nodata & \nodata & 24 \\
\hline
WASP-189b & \textit{Input:} & 14.2 \citetalias{WASP-189b1} & 0 & 0 & 7 & \nodata \\
{       } & \textit{Retrieved:} & $14.0^{+0.89}_{-0.86}$ & \nodata & \nodata & \nodata & 16\\
\hline
WASP-167b & \textit{Input:} & 10.7 \citetalias{WASP-167b1} & 11 & 88 & 5 & \nodata \\
{       } & \textit{Retrieved:} & $11.06^{+0.42}_{-0.67}$ & $11.1^{+0.72}_{-1.1}$ & $92.2^{+2.87}_{-8.88}$ & \nodata & 9\\
\hline
HAT-P-23b & \textit{Input:} & 8.2 \citetalias{HAT-P-23b} & 5 & 31 & 10 & \nodata \\
{       } & \textit{Retrieved:} & $7.5^{+0.40}_{-0.37}$ & $4.94^{+0.73}_{-0.74}$ & $34.5^{+3.11}_{-6.22}$ & \nodata & 6\\
\hline
HAT-P-70b & \textit{Input:} & 9.7 \citetalias{HAT-P-70b1} & 4.9 & 29 & 4 & \nodata \\
{       } & \textit{Retrieved:} & $10.6^{+0.63}_{-0.43}$ & $5.0^{+1.0}_{-1.1}$ & $30.6^{+2.1}_{-2.1}$ & \nodata & 5\\
\hline
Wendelstein-1b & \textit{Input:} & 5.8 \citetalias{Wendelsteinplanets} & 0 & 0 & 107 & \nodata \\
{       } & \textit{Retrieved:} & $5.4^{+0.23}_{-0.37}$ & \nodata & \nodata & \nodata & 4\\
\hline
WASP-103b & \textit{Input:} & 9.1 \citetalias{WASP-103b1} & 3.6 & 70 & 16 & \nodata\\
{       } & \textit{Retrieved:} & $10.1^{+0.74}_{-0.73}$ & $3.75^{+0.66}_{-0.94}$ & $98.6^{+20.0}_{-21.7}$ & \nodata & 4\\
\hline
HATS-24b & \textit{Input:} & 7.8 \citetalias{HATS-24b} & 4.9 & 31 & 12 & \nodata \\
{       } & \textit{Retrieved:} &  $8.54^{+0.67}_{-0.53}$ & $4.79^{+1.16}_{-0.75}$ & $32.8^{+2.59}_{-3.0}$ & \nodata & 4\\
\hline
WASP-178b & \textit{Input:} & 8.9 \citetalias{WASP-178b} & 0 & 0 & 3 & \nodata \\
{       } & \textit{Retrieved:} &  $9.04^{+0.38}_{-0.55}$ & \nodata & \nodata & \nodata & 4\\
\hline
Wendelstein-2b & \textit{Input:} & 5.5 \citetalias{Wendelsteinplanets} & 0 & 0 & 136 & \nodata \\
{       } & \textit{Retrieved:} &  $6.14^{+0.4}_{-0.35}$ & \nodata & \nodata & \nodata & 3\\
\hline
KELT-17b & \textit{Input:} & 10.5 \citetalias{KELT-17b2} & 0 & 0 & 2 & \nodata
\\
{       } & \textit{Retrieved:} & $10.1^{+0.36}_{-0.29}$ & \nodata & \nodata & \nodata & 3\\
\hline
HD 202882Ab & \textit{Input:} & 16.3 \citetalias{HD202772Ab} & 0 & 0 & 2 & \nodata \\
{       } & \textit{Retrieved:} & $16.9^{+0.89}_{-0.90}$ & \nodata & \nodata & \nodata & 3\\
\enddata
\tablerefs{1, \cite{HAT-P-23b}; 2, \cite{HAT-P-70b1}; 3, \cite{HAT-P-70b2}; 4, \cite{HATS-18b}; 5, \cite{HATS-24b}; 6, \cite{HD202772Ab}; 7, \cite{K2-260b}; 8, \cite{Qatar-7b}; 9, \cite{KELT-17b1}; 10, \cite{KELT-17b2}; 11, \cite{TOI-2109b}; 12, \cite{WASP-33b1}; 13, \cite{WASP-33b2}; 14, \cite{WASP-76b1}; 15, \cite{WASP-76b2}; 16, \cite{WASP-76b3}; 17, \cite{WASP-103b1}; 18, \cite{WASP-103b2}; 19, \cite{WASP-103b3}; 20, \cite{WASP-114b}; 21, \cite{WASP-167b1}; 22, \cite{WASP-167b2}; 23, \cite{WASP-178b}; 24, \cite{WASP-189b1}; 25, \cite{WASP-189b2}; 26, \cite{WASP189bTP}; 27, \cite{Wendelsteinplanets}}
\end{deluxetable*}

We first simulated the phase curve using our full phase curve model described in Equation \ref{eq:fullphasecurve}. For all of our simulations, we assume an $\alpha$ of 1, giving us a thermal linear gradient across the planet from the hottest slice to the coldest slice.

For each planet in our sample with a known day-side and night-side temperature, we set these temperatures to be the hemispherically-averaged temperature of the day-side and night-side, respectively. Because we assumed a linear thermal map of the planet, we can calculate the terminator temperature by taking the mean of the hemispherically-averaged day- and night-side temperature. Since thermal emission is distributed linearly across the planet, the hottest slice of the planet will be closest to the host star and the coldest slice will be farthest from the host star. We calculated the temperature of the hottest and coldest slice on the planet using the following equation,

\begin{equation}
\label{eq:max slice}
    T_{\text{max/min slice}} = (2*T_{\text{hemisphere}}) - T_{\text{terminator}}
\end{equation}

where $T_{\text{hemisphere}}$ is the day- or night-side temperature and $T_{\text{terminator}}$ is the terminator temperature. We calculated the blackbody radiation ratio of the coldest slice's temperature over the hottest slice's temperature and used this as the planet's $f_{drop}$ value.

For each planet in our sample with no known day-side or night-side temperature or both, we used the reported equilibrium temperature of the planet as the hemispherically-averaged day-side temperature. We estimated the mean $f_{drop}$ parameter from the planets in our sample with known day- and night-side temperatures to be about 0 ($\sim0.03$), giving us an $f_{thermal}$ of 1.0. We took the following steps to ensure the $f_{drop}$ parameter for our planets with no known day- or night-side temperature were $\sim$ 0. To calculate the temperature of the hottest slice of the day-side, we analytically calculated the $T_{\text{terminator}}$ using a range of potential night-side temperatures from 900 - 2300 K. This array of potential terminator temperatures was then used in Equation \ref{eq:max slice} to calculate the temperatures of the hottest and coldest slice on the planet. The night-side temperature and associated terminator temperature that produced an $f_{drop}$ value closest to 0.05 (which is $\sim$0) was used to find the temperature of the hottest slice in Equation \ref{eq:max slice}.

After simulating the phase curves, we added realistic photon noise to simulate data. We calculated the expected photon noise for each planet using the same method as the pure photon noise calculation in Section \ref{susbsubsec:noise}. Instead of stacking half of the orbital period as the exposure time, we used a cadence of 2 minutes as our exposure time. If the estimated photon noise was below the detector's noise floor, then the noise floor was used as the photon noise for the data simulation.

After estimating the noise, we use the \texttt{random.normal} function and set the simulated phase curve as the mean of the distribution and our noise level as the standard deviation. The random samples drawn from this function produced our mock data with simulated photon noise and represent a realistic data simulation of what \textit{Twinkle} will be able to achieve given our reported telescope and target parameters. After simulating the data, we then masked out every other 45 minutes of data to simulate the minimum 50 $\%$ overhead. 

\subsubsection{Spectroscopic Phase Curve Simulation} \label{subsubsec:spectroscopic phase curve sim}

We simulated spectroscopic phase curves for one of the planets in our sample: WASP-189b. We chose this planet as our case study because it has one of the highest SNRs in our sample and has been well-characterized in past publications (e.g. \citealt{WASP-189b1}, \citealt{WASP-189b2}). 

We first used \texttt{petitRADTRANS} to create an emission contribution function for WASP-189b to understand what atmospheric pressures and temperatures \textit{Twinkle} will probe \citep{petitradtrans}. We assumed a hydrogen and helium dominated atmosphere, typical of gas giant planets, with the following mass fractions and elements: $H_2$ - 0.74, $He$ - 0.24, $H_2O$ - $10^{-2}$, $CO$ - $10^{-4}$, $CO_2$ - $10^{-2}$, $K$ - $10^{-2}$, and $Na$ - $10^{-2}$. The mass fraction of water was consistent with what was used to simulate the measured abundance of water in WASP-18b, another ultra-hot Jupiter \citep{water_detection}. While the choice of abundances are arbitrary and for demonstration purposes, we provide the following cautionary notes. We note that WASP-18b does have a cooler equilibrium temperature than WASP-189b, and thus WASP-189b's water abundance could be lower due to dissociation. We assumed the same P-T profile as \cite{WASP189bTP} with a temperature inversion in its atmosphere (e.g. \citealt{WASP-189b2}, \citealt{WASP189bTP}). We also assume a reference gravity of $10^{3.5}$ m/s and a mean molar mass of 2.33. We then generated an emission contribution function under these assumptions.

We divided \textit{Twinkle's} wavelength range into thirty-two equal sized bins, which roughly correspond to the resolving power of 50-70. We found the weighted median of each wavelength bin from the emission contribution function and recorded the corresponding pressure at that mean. We then used these pressures in our P-T profile to find the corresponding temperature at that pressure level in the atmosphere. We used these temperatures as the planet temperature in our $B_{sp}$ parameter for each spectroscopic phase curve. We simulated the phase curve model and data using the same process as our photometric phase curves.

From the thirty-two simulated spectroscopic phase curves, we created phase-resolved emission spectra for WASP-189b at four phases. We first binned each spectroscopic phase curve into 15 equally sized bins to represent 15 phases. We then selected four different phases and plotted those points in relative flux versus the wavelength at which the phase curve is observed. The resulting phase-resolved emission spectra is shown in Figure \ref{fig:emissionspectrum}. Phase 0 represents the primary transit, phase 0.25 represents the crescent phase between the primary transit and secondary eclipse, phase 0.5 represents the secondary eclipse and phase 0.75 represents the crescent phase between the secondary eclipse and primary transit as described in Figure \ref{fig:modeldiagram}.

\subsection{Model Fitting} \label{subsec:model fitting}

We utilized the dynamic nested sampler \texttt{dynesty} to estimate Bayesian posteriors for all free parameters in both our photometric and spectroscopic phase curve fits \citep{dynesty_main}. \texttt{dynesty} will stop sampling when the log ratio between the current estimated evidence and the remaining evidence reaches a certain threshold. This threshold corresponds to the fraction of the evidence that remains unaccounted for. For our fits, we set this threshold to the default value of 0.01.

We allowed all parameters described in \ref{subsec:phase curve model} to vary freely in our model fits, and we set a uniform prior for the albedo amplitude, $\epsilon$, such that the lower and upper bounds could be converted into a geometric albedo between 0 and 1 using Equation \ref{eq:geometric albedo}. For flux drop, $f_{drop}$, we allowed it to vary between 0 and 1.5. The flux of the hottest slice on the planet is assigned a normalized value of 1 by design (see Equation \ref{eq:thermal}). Thus, the $f_{drop}$ parameter should only physically vary between 0 and 1. We've allowed it to vary between 0 and 1.5 because the hemispherically-averaged flux remains positive in the presence of un-physical negative flux. We set a uniform prior for the thermal phase offset, $\delta$, and allowed it to vary between the physically possible values of $\pi$ and $-\pi$. Finally, the interpolation power, $\alpha$, was allowed to vary between 0.5 and 2.0.

A Gaussian prior was used for the $R_{sp}$ as it can be reasonably constrained by transit observations. The mean for the prior distribution was calculated from the stellar and planetary radii from the sources in Tables \ref{tab:planetparameters1} and \ref{tab:planetparameters2}. The standard deviation for the prior distribution was based upon the maximum and minimum ratios calculated from the stellar and planetary radii reported in the NASA Exoplanet Archive. A Gaussian prior was used for $B_{sp}$ as both $R_{sp}$ and $B_{sp}$ contribute to the planet-star flux ratio as shown in equation (Eq. \ref{eq:thermal2}) in our model. The mean for the prior distribution was calculated from the stellar and planetary temperatures from the sources in Tables \ref{tab:planetparameters1} and \ref{tab:planetparameters2}. The standard deviation for the prior distribution was based upon the maximum and minimum ratios calculated from the stellar and planetary temperatures reported in the NASA Exoplanet Archive. For select planets that we modeled secondary effects, Gaussian priors were used for the Doppler boosting and ellipsoidal distortion amplitudes. The mean of the ellipsoidal distortion amplitude prior is the median of the possible range of values listed in the second column of Table \ref{tab:harmonics}. The standard deviation was set for each planet such that it encompassed the entire range of possible values reported in column 2 of Table \ref{tab:harmonics}. The mean of the Doppler boosting amplitude prior is the value reported in the third column of Table \ref{tab:harmonics}. The standard deviation was set for each planet such that it included all possible amplitude values calculated from reported parameters in the NASA Exoplanet Archive.

\section{Results} \label{sec:results}

\subsection{Photometric Phase Curves} \label{subsec:white light curves}

\begin{figure*}
\label{fig:photometric phase curves}
\gridline{\fig{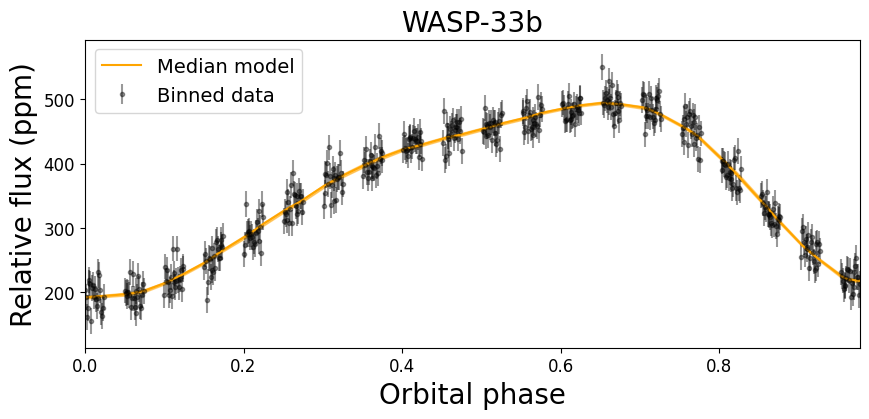}{0.33\textwidth}{(a)}
          \fig{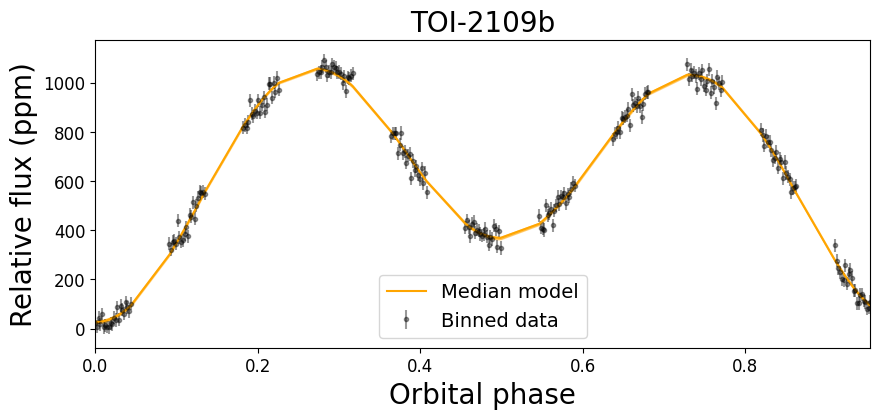}{0.33\textwidth}{(b)}
          \fig{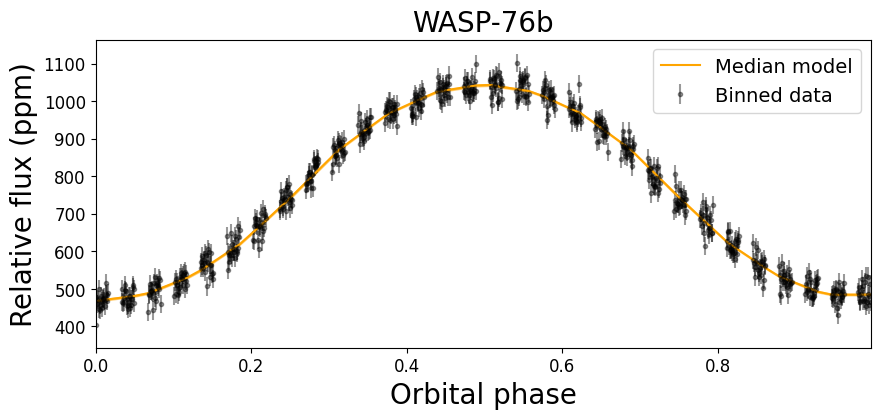}{0.33\textwidth}{(c)}}
\gridline{\fig{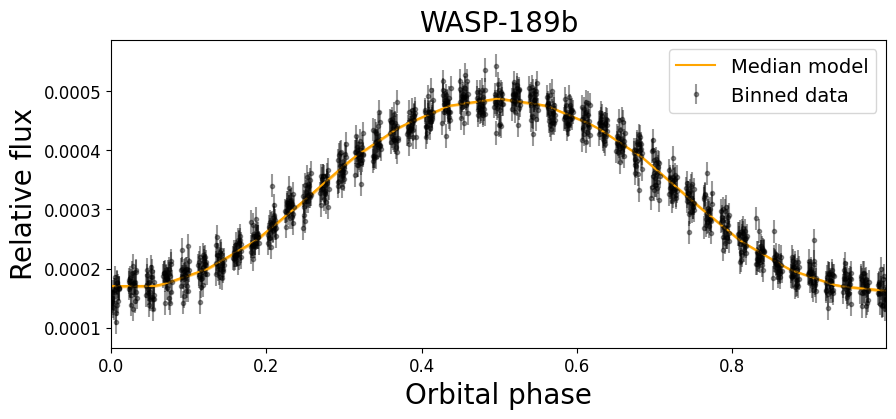}{0.33\textwidth}{(d)}
          \fig{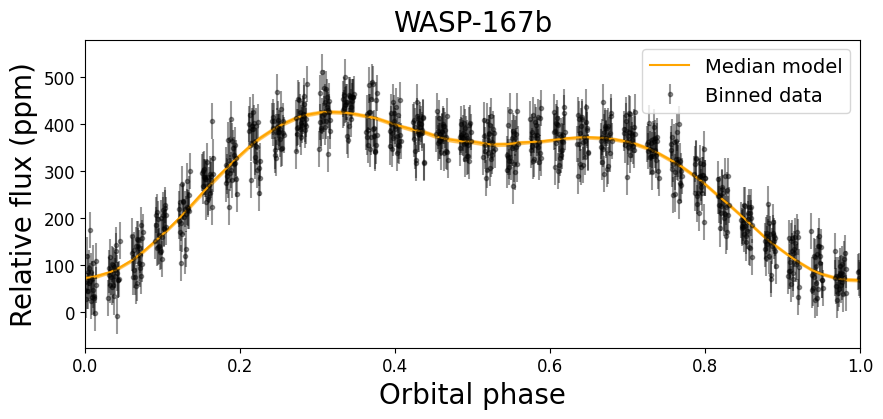}{0.33\textwidth}{(e)}
          \fig{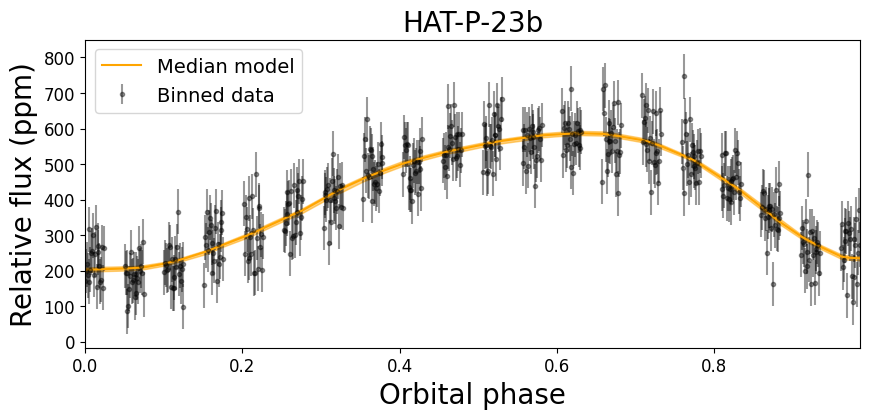}{0.33\textwidth}{(f)}}
\gridline{\fig{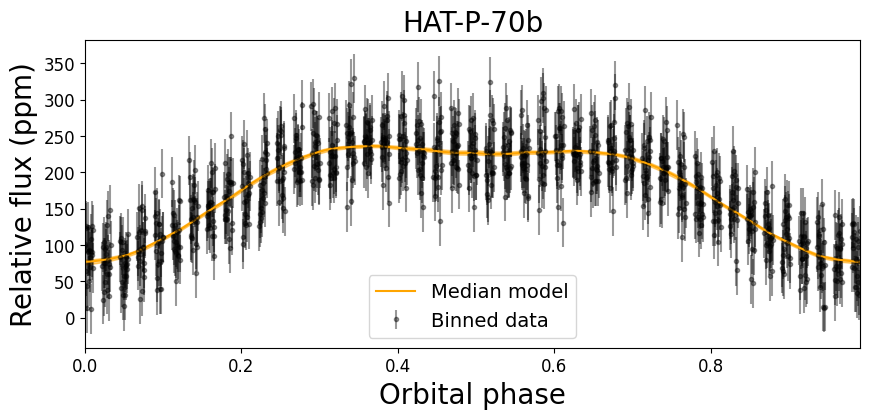}{0.33\textwidth}{(g)}
          \fig{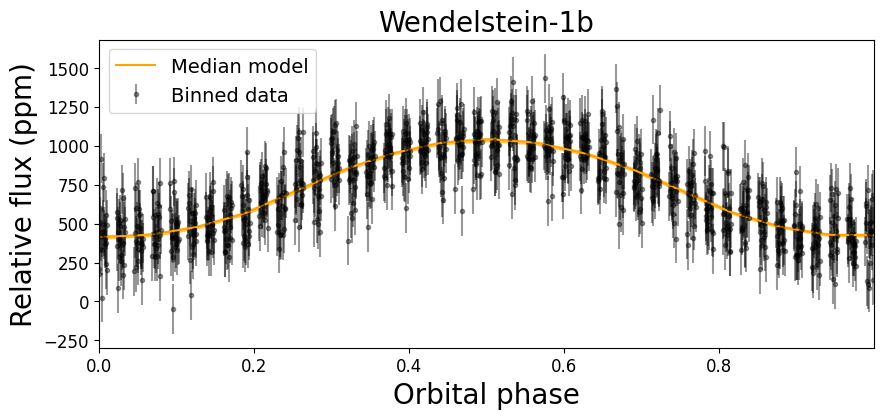}{0.33\textwidth}{(h)}
          \fig{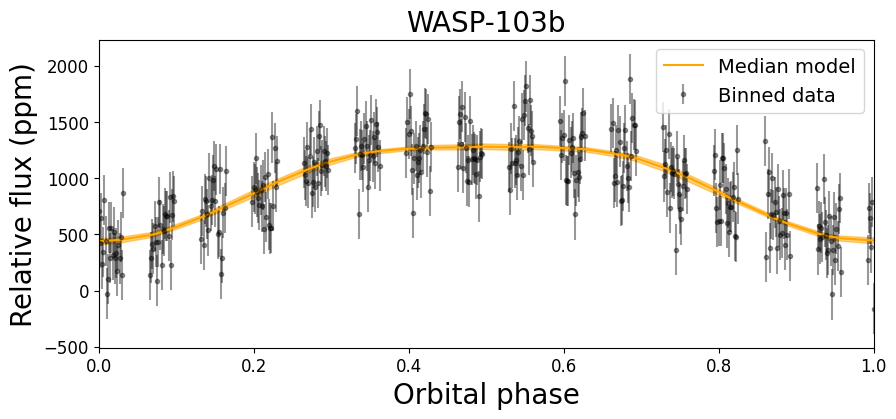}{0.33\textwidth}{(i)}}
\gridline{\fig{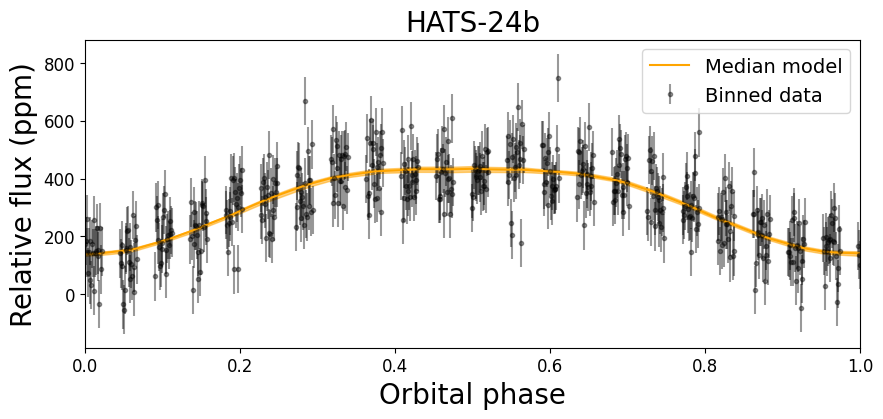}{0.33\textwidth}{(j)}
          \fig{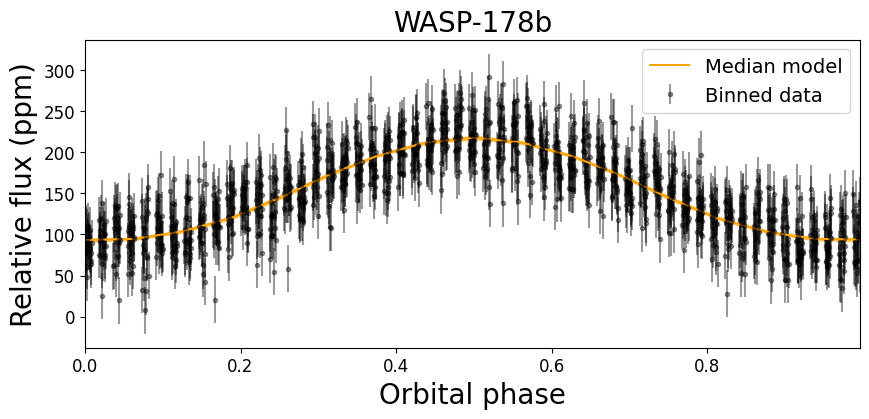}{0.33\textwidth}{(k)}
          \fig{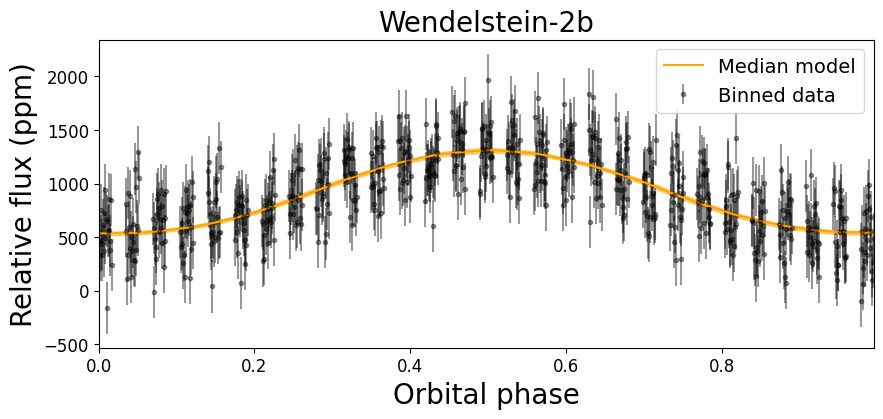}{0.33\textwidth}{(l)}}
\gridline{\fig{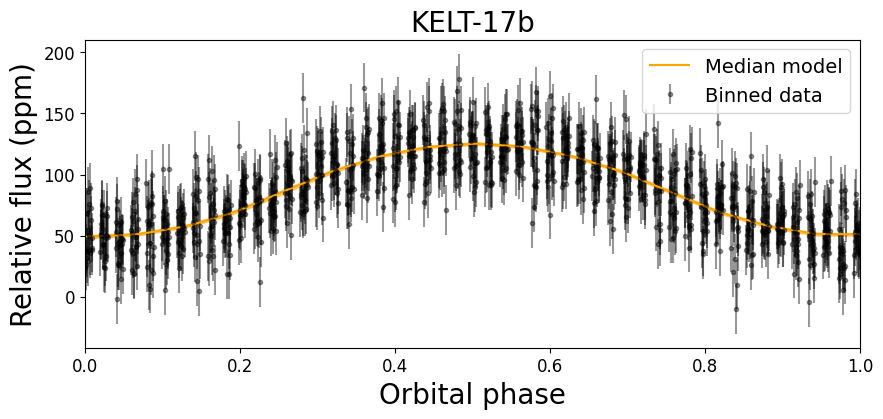}{0.33\textwidth}{(m)}
          \fig{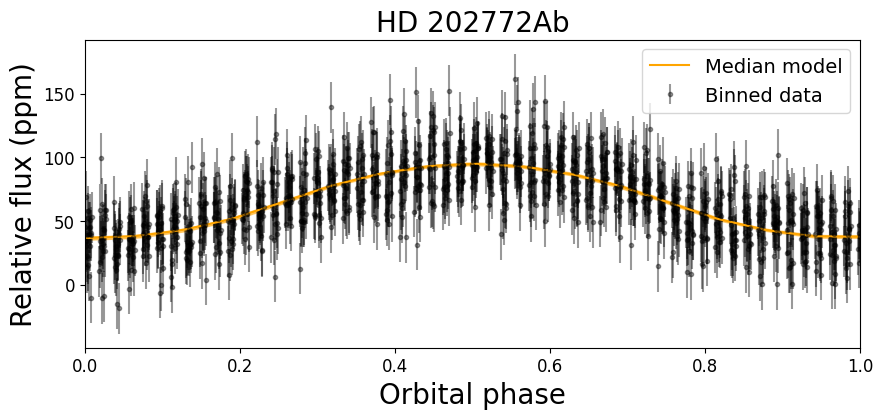}{0.33\textwidth}{(n)}}
\caption{Retrieved photometric phase curves of our survey with the \textit{Twinkle} satellite. The orange shaded region represents the 16th to 84th percentile of our \texttt{dynesty} run.}
\end{figure*}

Results of our retrieval for each planet along with our input parameters can be found in Tables (\ref{tab:planetparameters1} and (\ref{tab:planetparameters2}. We present results from our simulated photometric phase curve survey of UHJ atmospheres with \textit{Twinkle} in Figure \ref{fig:photometric phase curves}. Mock noisy data is in black and our best-fit model in orange. The orange shaded regions in each of the phase curves represent the 16th through 84th percentile results of our Bayesian posterior analysis. We retrieved a phase curve amplitude for all planets in our survey with a significance higher than at least 3$\sigma$, indicating that \textit{Twinkle} will be capable of detecting the phase curve signals for all planets in our survey. From the error bars on our retrieved phase offsets presented in Table \ref{tab:planetparameters1}, \textit{Twinkle} will be capable of detecting offsets in photometric phase curves within $\sim$0.5-3$\degree$. The satellite will be the first to observe the phase curves of over half of the planets in our sample.

For a small subsample of our planets, the modeled Doppler boosting and ellipsoidal distortion amplitudes affected the shape of the phase curve signal. These planets are TOI-2109b, WASP-167b, and HAT-P-70b. In the case of HAT-P-70b, no known phase curve observations have been taken to compare our theoretical ellipsoidal distortion and Doppler boosting amplitudes. We note that we used HAT-P-70b's upper mass constraint of ~6.78 $M_J$ in our calculations, which contributes to the double-peak structure in the phase curve \citep{HAT-P-70b1}.

In previous phase curve observations of TOI-2109b, the ellipsoidal distortion amplitude has dominated the shape of the phase curve signal \citep{TOI-2109b}. We find our theoretical ellipsoidal distortion amplitude and phase curve shape agree well with these past observations. In the case of WASP-167b, past phase curve observations have retrieved an ellipsoidal distortion signal that's one magnitude smaller than our theoretical value \citep{WASP-167b2}. However, stellar pulsations from WASP-167 have made measuring WASP-167b's mass difficult and led to discrepancies (\citealt{WASP-167b1}, \citealt{WASP-167b2}). For the purpose of our simulations, we use the upper mass constraint reported from \cite{WASP-167b1} to model the ellipsoidal distortion and Doppler boosting amplitudes. Although these effects can be modeled out of observations to retrieve the phase curve signal of the planet, they can still provide mass constraints from the measured mass ratio between the planet and star. With \textit{Twinkle}, we will be able to measure these secondary effects and provide a more comprehensive view of the star-planet system.

The retrieved parameters agreed within 1$\sigma$ of the input parameters for all of the planets in our sample. This indicates the phase curve offset, which measures the longitude of the brightest hemisphere, and the phase curve amplitude, which measures the day-night temperature contrast, can be retrieved from photometric phase curves observed by \textit{Twinkle}. Putting constraints on these parameters will constrain other global properties of planetary atmospheres and allow us to draw conclusions about their dynamics and composition. We discuss this in detail in Section \ref{sec:discussion}.

\subsection{Spectroscopic Phase Curves} \label{subsec:specphasecurves}

In addition to photometry, our survey will make use of \textit{Twinkle}'s spectroscopic capabilities to probe UHJ atmospheres across optical and infrared wavelengths. Spectroscopic phase curves can be used to probe planetary properties and dynamics by wavelength and provide more detail than photometric phase curves can. In our planned survey, we will observe both photometric and spectroscopic phase curves for all of our planets. For the purpose of this study, we simulated spectroscopic phase curves of one of our targets, WASP-189b \citep{WASP189discovery}. We selected WASP-189b as our case study because it has been the subject of numerous studies since it's discovery due to it's host star's brightness and the planet's large signal (e.g. \citealt{WASP189bparameters}, \citealt{WASP189study1}, \citealt{WASP189study2}). The planet's phase curve was observed in the optical with the \textit{CHEOPS} satellite, however no phase curve observations for the planet yet exist in the infrared \citep{WASP189bparameters}. This presents a unique opportunity for \textit{Twinkle} to observe WASP-189b simultaneously in the optical and infrared wavelengths, providing a more complete characterization of the planet's atmosphere.

\begin{figure*}[th!]
\centering
\includegraphics[width=0.95\textwidth]{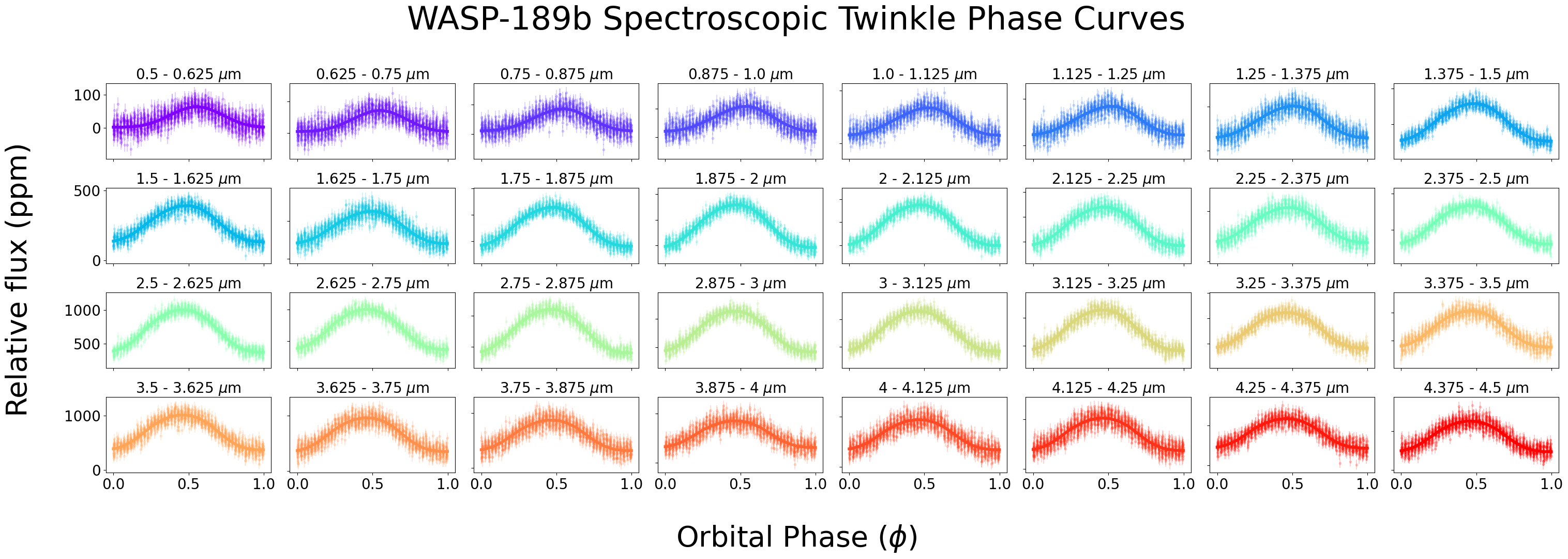}
\caption{Simulated spectroscopic phase curves for WASP-189b in 32 wavelength bins within \textit{Twinkle}'s wavelength range. The y-axis is the relative flux in parts-per-million and the x-axis is orbital phase in radians. The wavelength range for each bin is indicated above each phase curve. 
\label{fig:specphasecurves}}
\end{figure*}

We simulated spectroscopic phase curves for WASP-189b from 0.5 to 4.5 $\micron$ shown in Figure \ref{fig:specphasecurves}. Our process for simulating the phase curves is described in Section \ref{subsubsec:spectroscopic phase curve sim}. We used the same parameters as the photometric phase curves in our simulated spectroscopic observations. WASP-189b has an elevated geometric albedo of 0.26 which provides an ample reflected light component probed in the first four phase curves in Figure \ref{fig:specphasecurves} \citep{WASP-189b2}. At longer wavelengths, the thermal emission of the planet dominates the phase curve observations, causing the amplitude to increase and allowing us to better examine the day- and night-side temperature contrast.

Spectroscopic phase curves allow us to directly probe the reflection and thermal components of phase curves by dividing observations by wavelength. Reflected light from exoplanets dominates in optical wavelengths, while the thermal emission from the planet itself dominates in thermal wavelengths (e.g. \citealt{thermal_component}, \citealt{reflected_component}). Phase offsets, if present, can be either eastward or westward. Eastward phase offsets indicate the brightest, and therefore hottest, point on the planet has been offset to the east of the substellar point. Thermal wavelengths will best probe this type of phase offset if it's due to thermal emission. Westward phase offsets may indicate the presence of clouds among other phenomena (\citealt{westwardoffset}, \citealt{magneticfield}). Optical wavelengths will best probe this type of phase offset if it's due to cloud formation on the western hemisphere. For more details on these processes, see Sections \ref{subsec:westwardoffset} and \ref{subsec:jetstructure}.

The phase offset is a free parameter in the thermal component of our full phase curve model. We manually set the phase offset in the reflection component of our model to $\pi$ for all of our previous simulations. This is to ensure the reflected light component peaks at the secondary eclipse, or on the day-side of the planet. With spectroscopic phase curves, it is possible to distinguish different phase offsets in reflected light and thermal wavelengths.

To understand if our full phase curve model could distinguish between two different phase offsets at different wavelengths, we allowed the phase offset in the reflection component of the model to vary freely between 0 and 2$\pi$. See Sections \ref{subsec:westwardoffset} and \ref{subsec:jetstructure} for a discussion on what different offsets in different wavelengths We then simulated a 20 degree westward phase offset in the reflected light component and a 20 degree eastward phase offset in the thermal emission component for the spectroscopic phase curves of WASP-189b. Our 20 degree input value is consistent with the typical observed value (e.g. \citealt{Beatty2019}, \citealt{WASP-167b1}, \citealt{Dang2024}).

\begin{figure*}[th!]
\centering
\includegraphics[width=0.7\textwidth]{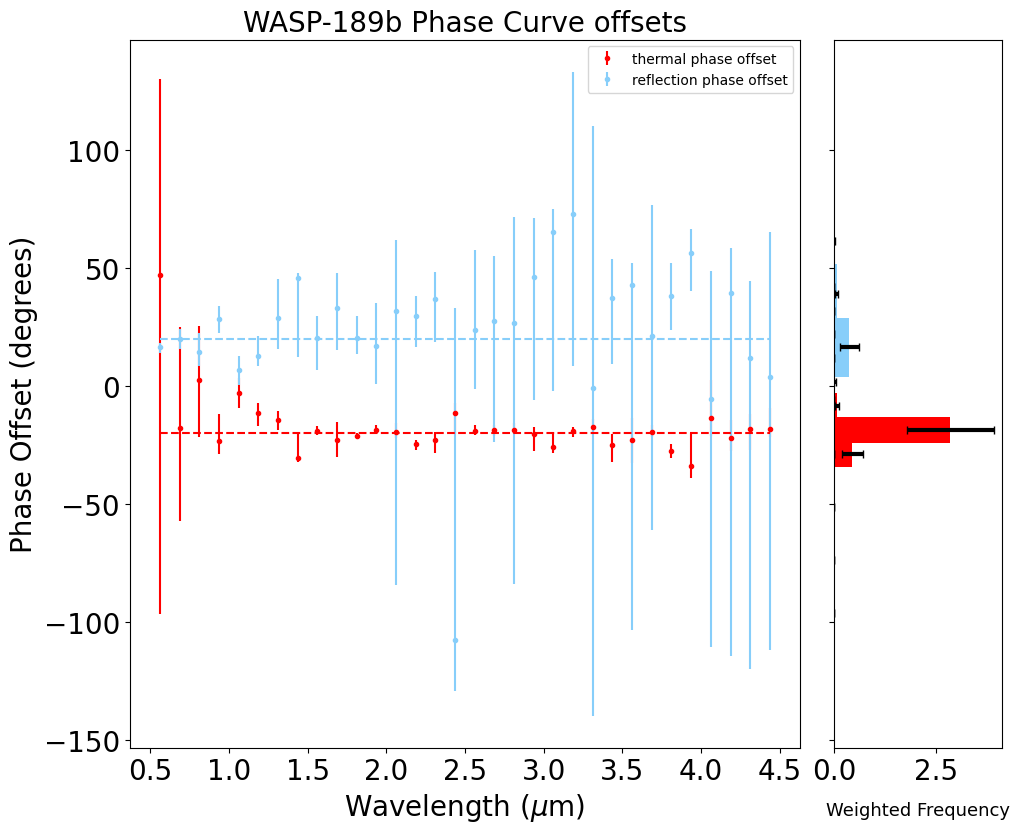}
\caption{Retrieved phase offsets from both reflection and thermal components. The y-axis is phase offset in degrees and the x-axis is wavelength in microns. The blue and red dotted lines represent the truth value of the phase offsets in the  reflection and thermal components, respectively. The histogram on the right shows the frequency of the points scaled by weight around the truth value for each component.
\label{fig:phase shifts}}
\end{figure*}

We retrieved the Bayesian posteriors of the phase offsets in the different components and the results are shown in Figure \ref{fig:phase shifts}. The upper error bar and lower error bar of each point is the 16th and 84th percentile of the Bayesian posteriors, respectively. The simulated 20 degree westward offset in the reflection component was constrained in optical wavelengths and became unconstrained in the thermal wavelengths. Vice versa, the thermal phase offset remained less constrained in the optical wavelengths and became more constrained in the thermal wavelengths. Since the error bars are unequal on most of our points, we took the average of the upper and lower error bars for each point. This average was then used to calculate the weight that each point would contribute to the overall bin count. The weight was calculated by taking $\frac{1}{avg^2}$ where $avg$ is the average of the upper and lower error bars. We binned the 32 retrieved phase offsets in both components into 8 bins and created a histogram using \texttt{Matplotlib's plt.hist()} function. We chose 8 bins for our histogram to reveal a Gaussian-like distribution of the points while also keeping the bin count as low as possible. We found a sharp peak of -20.2$\degree$ with a standard deviation of 0.5$\degree$ for the thermal phase offset and a smaller peak at 19.0$\degree$ with a standard deviation of 1.5$\degree$ for the reflected light phase offset. \textit{Twinkle} therefore will be able to distinguish between phase offsets in the optical and infrared wavelengths, with the infrared wavelengths more sensitive to phase offset measurements. 

\begin{figure*}[th!]
\centering
\includegraphics[width=0.8\textwidth]{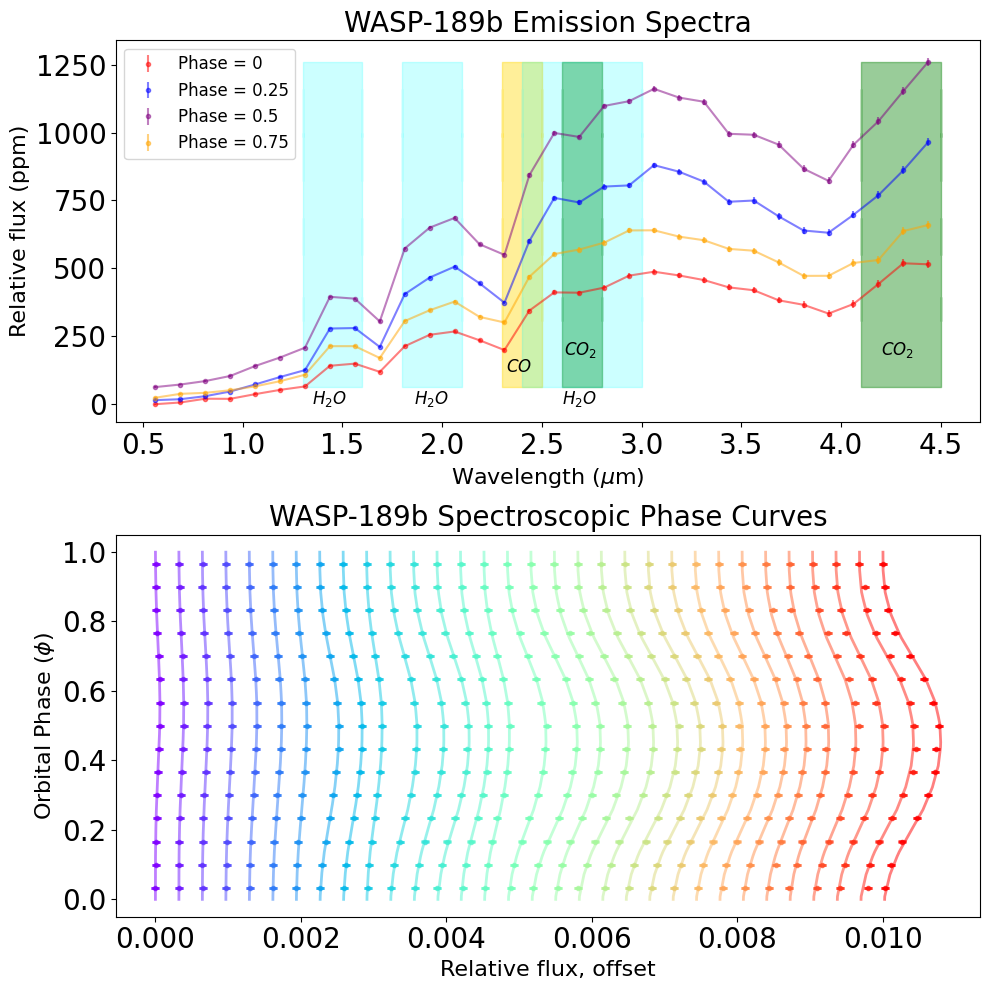}
\caption{\textit{Top panel}: Phase-resolved emission spectra of WASP-189b at the selected phases 0, 0.23, 0.5 and 0.78. The broad molecular features within \textit{Twinkle's} wavelength range are highlighted underneath the spectra. For clarity, we only highlight the molecules used in the simulation of WASP-189b's atmosphere. \textit{Bottom panel}: the spectroscopic phase curves shown in Figure \ref{fig:specphasecurves} binned into 15 phases. These binned phase curves were used to create the phase-resolved emission spectra in the top panel. The color of each binned phase curve correspond to the same phase curve in Figure \ref{fig:specphasecurves}.
\label{fig:emissionspectrum}}
\end{figure*}

Previous studies of spectroscopic phase curves have constructed phase-resolved emission spectra (e.g. \citealt{HSTspecphasecurves}). Using our spectroscopic phase curves, we create emission spectra for WASP-189b at four selected phases. The binned spectroscopic phase curves used to create the spectra are shown in Figure \ref{fig:emissionspectrum}. This process is detailed in Section \ref{subsubsec:spectroscopic phase curve sim}. We did not calculate atmospheric chemistry when modeling the emission spectra. As described in Section \ref{subsubsec:spectroscopic phase curve sim}, we see broad molecular features such as water, carbon dioxide and carbon monoxide in the emission spectra. We note that WASP-189b may not have water in its atmosphere due to its high equilibrium temperature of 3353 K. We provide this simulation as a case study only to demonstrate what mass fractions of molecules \textit{Twinkle} will be sensitive to. Phase-resolved emission spectra like this can be used to map the planet's changing chemistry across its hemispheres. For more possible science cases using the phase-resolved emission spectra, see \ref{subsec:moleculedetection}. We do not run a spectral retrieval on the data as its beyond the scope of this paper, but this simulation represents \textit{Twinkle's} ability to construct emission spectra out of phase curve observations.

\subsection{Spherical Geometry Correction} \label{subsec:spherical geometry corr}

\begin{figure*}[th!]
\centering
\includegraphics[width=0.8\textwidth]{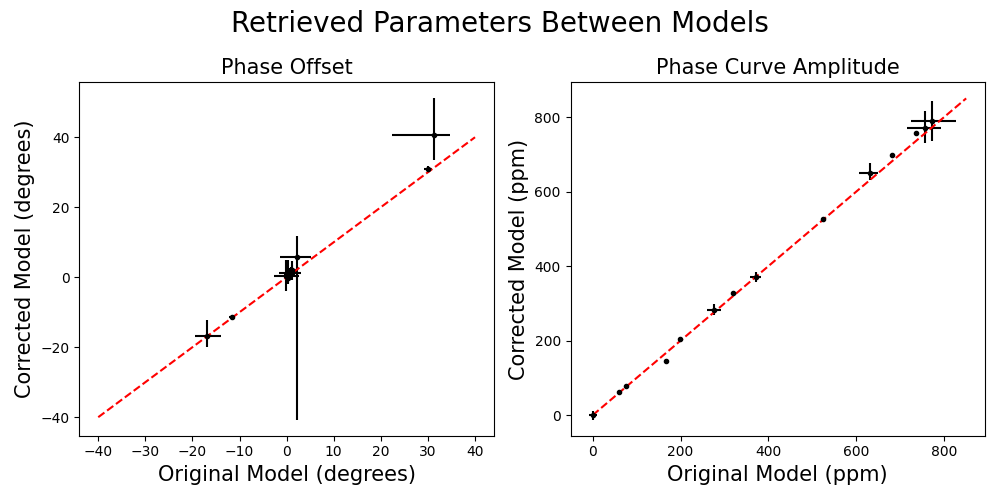}
\caption{Comparison between geometry-corrected model and original phase curve model. Each black point is represents one of the planets in our survey. The red-dashed line in both is a line of equality.
\label{fig:modelcorrection}}
\end{figure*}

We modeled the thermal emission of each planet in our survey by slicing the planet into sixteen longitudinal slices as described in Section \ref{susbsubsec:thermal}. Each slice in our full phase curve model is assumed to have equal area and provide equal contribution to the thermal contribution at each phase. However, from spherical geometry, the projected area of each slice will vary and will physically not have equal weight. We introduced a correction factor to the thermal contribution (Equation \ref{eq:thermal}) of each slice in order to see if this affected our results. We did not include a correction factor to the reflected light contribution because Equation \ref{eq:reflection} is already phase corrected. We used the following equation as our correction factor,

\begin{equation}
\label{eq:corrected thermal}
    \sum_{i = 0}^{M} f_{thermal, i} \times \cos({\theta_i})
\end{equation}

where $f_{thermal, i}$ is Equation \ref{eq:thermal} at each visible slice i at a given phase and $\theta_i$ is the angle of each visible slice i with respect to the line of sight. We defined this angle as,

\begin{equation}
    \theta_i = -90\degree + \frac{180 \degree}{M+1} i
\end{equation}

Thus, Equation \ref{eq:corrected thermal} replaces Equation \ref{eq:thermal} in our full phase curve model (\ref{eq:fullphasecurve}).

We then ran the corrected model on each of our photometric phase curves using the same process as the original model (see Section \ref{subsec:phase curve model} for more details). We compared the retrieved values of the thermal phase curve offset and phase curve amplitude between both models. This comparison is shown in Figure \ref{fig:modelcorrection}. For both parameters, the values from the original and corrected models are consistent within their respective error bars, showing no change in the effectiveness of our model.

\section{Science Cases} \label{sec:discussion}

In this section, we explore possible science cases with our photometric and spectroscopic phase curves. We have shown that phase curve offsets of UHJs will be detectable within $\sim$ 0.5 - 3$\degree$ in photometric phase curves and within $\sim$ 0.5$\degree$ in thermal and $\sim$ 1.5$\degree$ in reflected light spectroscopic phase curves with \textit{Twinkle}. We discuss what information can be inferred from the possible detection of westward offsets in Subsection \ref{subsec:westwardoffset} and eastward offsets in Subsection \ref{subsec:jetstructure}. As shown in Figure \ref{fig:emissionspectrum}, broad molecular features of CO, CO$_2$ and H$_2$O will be visible in phase-resolved emisson spectra. We discuss the significance and potential implications of molecular detections in Subsection \ref{subsec:moleculedetection}. Finally, we discuss any potential trends we could observe in UHJ atmospheres with our phase curve survey in Subsection \ref{subsec:populationlevel}.

\subsection{Presence of clouds, magnetic fields and non-synchronous rotation in UHJ atmospheres}
\label{subsec:westwardoffset}
The dayside temperatures of UHJs are unlikely to host clouds due to their high temperatures \citep{HJclouds}. However, there is evidence to support that the night-side temperatures could be low enough to allow clouds to form \citep{UHJnightsideclouds}. It has been theorized that the presence of clouds could potentially produce a westward phase offset in a phase curve \citep{HJclouds}. This type of phase offset was first detected on CoRoT-2b using the \textit{Spitzer} Space Telescope. Since then, westward phase offsets have been detected for a few ultra-hot Jupiters (e.g. \citealt{WASP33bparameters}, \citealt{WASP121boffset}, \citealt{westwardoffset}). If the western side of an UHJ's day-side is cold enough, it could allow for clouds to form near the cooler side of the terminator. Clouds are reflective, so this would induce a westward peak in the phase curve that would be most sensitive in optical wavelengths, right within \textit{Twinkle's} wavelength range.

Westward offsets could also be due to other factors such as magnetic effects and non-synchronous rotation. In hydrodynamic simulations of UHJ atmospheres, an eastward phase offset is predicted due to the simulated magnitude of winds \citep{easternhotspot}. These winds have shown to be variable in the presence of a significant magnetic field, shifting from flowing east to west and instead flowing from north to south. This effect could potentially cause a westward phase offset as the shift in wind flow causes heat to build on the western side of the night-side hemisphere \citep{magneticfield}. Additionally, in the absence of east-to-west flowing winds, the temperature contrast of the day- and night-side will increase. Probing this contrast at different thermal wavelengths, coupled with a westward offset, could help distinguish magnetic effects from other mechanisms of westward offsets.

The dominant underlying assumption in UHJ phase curves is that UHJs are tidally locked and therefore have a synchronous rotation rate with their orbital period. However, some theoretical studies have shown this assumption may not be accurate (e.g. \citealt{ArrasandSocrates}, \citealt{Socrates2012}, \citealt{slowrotation}). In that case, simulations of ultra-hot Jupiters have shown that with slower rotation rates, the atmospheres of these worlds may have a different temperature structure near their poles, resulting in large, cold vortices \citep{slowrotation}. Given the orientation of the planet towards the observer, these colder structures around the poles could potentially push the peak of the latitudinally average temperature towards the westward part of the planet. This could produce an observable westward phase offset \citep{slowrotation}. Additionally, if this phase offset does not change across wavelength (i.e. spectroscopic phase curves), then this could be evidence of non-synchronous rotation as well \citep{Showman2014}.

We have shown that \textit{Twinkle} will be able to detect offsets throughout its wavelength range, and in both reflection and thermal regimes. A potential westward offset produced by any of these mechanisms would be observable in our survey. The implications of even one westward offset observed in our sample would provide an interesting insight into the atmospheric dynamics of UHJs.

\subsection{Vertical Structure of Equatorial Jets}
\label{subsec:jetstructure}

Phase offsets occur when the longitude of the planet's brightest hemisphere is offset from the planet's substellar point. If a planet has an eastward phase shift, the presence of an equatorial jet can be inferred \citep{crossfield2018}. This is due to the energy and matter transport from the dayside to the nightside of tidally locked planets. This transport process can be characterized by three different timescales: the radiative, drag and gravity wave timescales. The amount of time it takes for gas to radiate their energy away is the radiative timescale. The drag timescale is the amount of time it takes for atmospheric winds to lose their kinetic energy and notably slow down. The gravity wave timescale is the characteristic time that a gravity wave needs to travel over the planet's radius. An eastward equatorial jet, and therefore an eastward phase offset, occurs when the radiative and drag timescales are longer than the gravity wave timescale \citep{crossfield2018}.

Phase curves observed at different wavelengths can offer different insights into the planet's atmosphere and probe their 2D thermal and chemical structure. Different wavelengths are sensitive to different molecular absorption bands, thus providing a proxy for the atmospheric depth at which that wavelength is observing \citep{crossfield2018}. Eastward phase offsets \citep{Knutson2007} observed at different wavelengths in spectroscopic phase curves would indicate that there is an equatorial jet at that atmospheric depth. With multiple phase curves observed at multiple wavelengths, the presence of equatorial jets at different atmospheric depths can be inferred. We've demonstrated that \textit{Twinkle} will be able to detect different phase offsets at different wavelengths, as shown in Figure \ref{fig:phase shifts}. This would provide additional characterization of the vertical structure of the planet as well as constrain longitudinal mixing at those depths based upon the magnitude of the phase offset. The larger the phase offset, the stronger the equatorial jet and strength of sideways mixing at those atmospheric depths.

\subsection{Detection of Molecules}
\label{subsec:moleculedetection}

As shown in Section \ref{subsec:specphasecurves}, phase-resolved emission spectra can be constructed from spectroscopic phase curves. While \textit{Twinkle's} spectral resolution is low (R = 50 - 70), broad molecular features will still be detectable. With phase-resolved emission spectra, we can map the changing chemistry across hemispheres. In Figure \ref{fig:emissionspectrum}, molecular features from water and carbon dioxide change in strength due to the changing temperatures across both sides. We can map these changing molecular abundances and potentially determine the life cycle of such molecules on these exotic worlds. 

Additionally, the presence of carbon- and oxygen-bearing molecules, such as CO, could be used to constrain a planet's metallicity relative to its host star \citep{metallicitycalc}. If the metallicity is subsolar, it can be inferred that the planet was formed outside the snow line of the system while a supersolar metallicity implies the planet migrated through the disk where metals are more present in the oxygen-rich planetesimals \citep{metallicitycalc2}. Similarly, detections of oxide molecules such as titanium oxide could strengthen the argument for potential clouds on the colder night-sides of UHJs \citep{oxideclouds}. Oxides such as titanium oxide have been proposed as tracers of a thermal inversion in the planet's atmosphere and thus, disequilibrium chemistry (e.g. \citealt{thermalinversion2}, \citealt{thermalinversion}). Titanium oxide has a number of molecular features within \textit{Twinkle's} wavelength range, particularly between 0.6 and 0.8 $\mu m$. This would provide insight into the thermal structure of UHJs, potentially revealing trends in thermal inversions in hot gas giant atmospheres. 

\subsection{Population Level Trends}
\label{subsec:populationlevel}

Population studies of UHJ phase curves have been done before (e.g. \citealt{spitzerphasecurvestudy}, \citealt{Wong2021}) using observations from \textit{Spitzer}, \textit{TESS}, \textit{Kepler} and \textit{Hubble}. However, there has not been a UHJ phase curve survey conducted using an instrument that simultaneously probes both the optical and infrared wavelength ranges. As mentioned before, the optical and infrared wavelengths both probe different, but vital, parts of an UHJ. Thermal phase curve surveys with \textit{Spitzer} revealed a positive correlation between incident flux and day-side temperatures of UHJs and optical phase curve surveys with \textit{TESS} revealed a positive correlation between geometric albedo and day-side temperature (\citealt{spitzerphasecurvestudy}, \citealt{Wong2021}). However, trends in other parameters, such as phase offset are more uncertain given large uncertainties and variations in de-trending models. With a consistent sample of UHJ phase curve observations in both the optical and infrared, degeneracies between the reflected light and thermal components can be broken. Additionally, consistent use of 1 de-trending model across the survey will allow us to more robustly and uniformly detect these trends.

Additionally, we've shown that our model can distinguish between phase offsets in the optical and infrared portions of \textit{Twinkle}'s wavelength range. Using our model, we can reveal potential trends in these offsets. A positive or negative correlation between optical offsets and differing planetary parameters (e.g. day-side and night-side temperature, planetary mass, orbital period) could potentially reveal the optimal conditions for cloud formation on UHJs. A positive or negative correlation between thermal offsets and different planetary parameters could potentially reveal optimal conditions for winds and atmospheric circulation on UHJs. Phase-resolved emission spectra from spectroscopic phase curves could reveal potential trends in elemental abundances that could impact planetary formation theories and inform general circulation models. By increasing the sample size of UHJs in a phase curve survey with \textit{Twinkle}, we will also be able to more robustly detect these trends.

\section{Conclusions} \label{sec:conclusions}
We present a simulation of a potential survey of ultra-hot Jupiter (UHJ) atmospheres via phase curves using the upcoming \textit{Twinkle} satellite. The following points are our primary conclusions:

\begin{itemize}
  \item Photometric and spectroscopic phase curve signals of the 14 UHJs included in our survey will be detectable by \textit{Twinkle} at minimum to 3$\sigma$ (Section \ref{fig:photometric phase curves}). This includes about half of our sample that has no known previous phase curve observations. Their amplitudes will reveal the day- and night-side temperature contrast of the planet.
  \item Phase offsets in photometric phase curves will be detectable within $\sim$ 0.5 - 3$\degree$ with \textit{Twinkle} (Section \ref{subsec:white light curves}). Additionally, \textit{Twinkle} spectroscopic phase curves will be able to distinguish between phase offsets in the reflected light (within $\sim$ 1.5$\degree$) and thermal emission (within $\sim$ 0.5$\degree$) of the planet (Section \ref{subsec:specphasecurves}). Offsets in different wavelengths will probe different aspects of a planet's atmosphere.
  \item Phase-resolved emission spectra can be constructed from spectroscopic phase curves and broad molecular features of CO, CO$_2$, and H$_2$O will be visible (Section \ref{subsec:specphasecurves}). Retrieving abundances of these molecules will inform planetary formation and evolution models.
  \item Potential new trends in planetary parameters such as phase offsets, albedos and incident flux will be revealed in our UHJ phase curve survey given the simultaneous coverage of optical and infrared wavelengths (Section \ref{sec:discussion}).
\end{itemize}

\section{Acknowledgments}
\begin{acknowledgments}
This research has made use of the NASA Exoplanet Archive, which is operated by the California Institute of Technology, under contract with the National Aeronautics and Space Administration under the Exoplanet Exploration Program. K.G. is grateful for the helpful discussions from Benjamin Wilcock and the \textit{Twinkle} Science Team. K.G. acknowledges support from NASA under Grant No. 80NSSC24K0365.
\end{acknowledgments}


\software{astropy (\citealt{astropy:2013}, \citealt{astropy:2018}, \citealt{astropy:2022}), petitRADTRANS (\citealt{petitradtrans}), dynesty (\citealt{dynesty_main}), matplotlib (\citealt{matplotlib})}



\appendix

\section{Model Validation}

Before data simulation, we tested our phase curve model against known TESS data for three different planets. These planets were chosen based upon the availability of TESS phase curve data and the reliability of previous analyses to compare to. The results of these tests compared to their literature values is reported in Table \ref{tab:modeltesting}. For all of our test planets, we utilized the same process for estimating Bayesian posteriors as described in Section \ref{subsec:model fitting}. Priors for each model fit and parameter can be found in Table \ref{tab:testingpriors}.

In the case of all three of our fits to the planets' TESS data, we used our full phase curve model, specified in Eq. \ref{eq:fullphasecurve}, with one variation. In each planet's case, we replaced the $f_{harmonics}$ part of Eq. \ref{eq:fullphasecurve} with the corresponding harmonics equation used in the literature for that specific planet. We specify the specific equation under each planet's section. We do this to ensure that we're modeling every effect that was present in the original analysis and to test the validity of the analytical parts of our model: the thermal and reflected light components. Every other part of our model described in Section \ref{subsec:phase curve model} remains the same.

The photon noise parameter reported in Table \ref{tab:modeltesting} was calculated using the same process as Section \ref{susbsubsec:noise}. This was to ensure our photon noise calculator was able to accurately reproduce the photon noise in real observations. All of our photon noise calculations were similar to the corresponding literature values.

\begin{deluxetable*}{lccccBcccccBcccc}
\tabletypesize{\scriptsize}
\tablewidth{0pt} 
\renewcommand{\arraystretch}{1.7}
\tablecaption{Parameter Results from Model Testing Compared to Literature Values \label{tab:modeltesting}}
\tablehead{
\colhead{Parameter} & \colhead{} & \colhead{KELT-9b}& \colhead{WASP-19b} & \colhead{Kepler-13Ab} &
} 
\startdata 
{Albedo}& \textit{Model:} & $0.14^{+0.08}_{-0.08}$ & $0.27^{+0.03}_{-0.03}$ & $0.74^{+0.07}_{-0.07}$ \\
{       }& \textit{Literature:} & 0 & $0.16^{+0.04}_{-0.04}$ & $0.53^{+0.15}_{-0.15}$ \\
\hline
{Amplitude (ppm)}& \textit{Model:} & $570^{+12}_{-12}$ & $543^{+85}_{-85}$ & $288^{+32}_{-34}$ \\
{       }& \textit{Literature:} & $566^{+16}_{-16}$ & $626^{+100}_{-100}$ & $302^{+30}_{-32}$\\
\hline
{Phase offset (degrees)}& \textit{Model:} & $5.7^{+0.8}_{-1}$ & $1.7^{+103}_{-103}$ & $60^{+69}_{-62}$\\
{       }& \textit{Literature:} & $5.2^{+0.9}_{-0.9}$ & 0 & $8.9^{+5.0}_{-4.6}$\\
\hline
{Doppler boosting amplitude (ppm)} & \textit{Model:} & $A_1$ = $2.1^{+0.3}_{-0.3}$ & $47^{+27}_{-25}$ &  
$7.1^{+1.6}_{-1.6}$\\
{         }& {\textit{Literature:}} & $A_1$ = $2.1^{+0.3}_{-0.3}$ & $49^{+45}_{-44}$ & $6.8^{+1.7}_{-1.7}$ \\
\hline
{Ellipsoidal distortion amplitude (ppm)}& \textit{Model:} & $B_2$ = $6.1^{+3.3}_{-1.4}$& $0^{+32}_{-26}$  & $58^{+6}_{-9.5}$ \\
{     }& {       } & $B_3$ = $-13^{+3.1}_{-2.0}$ & \nodata  &  \nodata \\
{       }& \textit{Literature:} & $B_2$ = $-16.1^{+6.0}_{-5.9}$ & $12^{+51}_{-48}$ & $49^{+17}_{-16}$ \\
{       } & & $B_3$ = $-3.0^{+6.4}_{-6.2}$ & \nodata &  \nodata \\
\hline
{Systematics and other effects (ppm)}& \textit{Model:} & $A_2$ = $-42^{+2}_{-1.2}$ & $-84^{+23}_{-23}$ & \nodata \\
{       }&  & $A_3$ = $14^{+2.7}_{-2.8}$ & \nodata  &   \nodata\\
{       }& \textit{Literature:} & $A_2$ = $-35.7^{+4.2}_{-4.3}$ & $-86^{+38}_{-38}$  &     \nodata \\
{       }& {       } & $A_3$ = $13.9^{+4.2}_{-4.3}$ & \nodata &     \nodata \\
\hline
{Photon noise (ppm)}& \textit{Model:} & 28 & 223 & 73\\
{       }& \textit{Literature:} & 25 & 300 & 75\\
\hline
{Reference}&  & \citep{Wong2020kelt9} & \citep{Wong2020wasp19} & \citep{Wong2021} \\
\enddata
\tablecomments{List of all free parameters in the modeling test of three planets using our full phase curve model and the literature values to compare.}
\end{deluxetable*}

\begin{deluxetable*}{lccccBcccccBcccc}
\tabletypesize{\scriptsize}
\tablewidth{0pt} 
\tablecaption{Priors for Model Fits to TESS Data \label{tab:testingpriors}}
\tablehead{
\colhead{Parameter} & \colhead{Units} & \colhead{KELT-9 b}& \colhead{WASP-19 b} & \colhead{Kepler-13A b}
} 
\startdata 
Albedo amplitude & { } & $\mathcal{U}(0,\,0.00068)$ & $\mathcal{U}(0,\,0.00168)$ & $\mathcal{U}(0,\,0.00036)$ \\[1mm]
Flux drop & { } & $\mathcal{U}(0,\,1.4)$ & $\mathcal{U}(0,\,1.4)$ & $\mathcal{U}(0,\,1.4)$ \\[1mm]
Thermal Offset & radians & $\mathcal{U}(-\pi,\,\pi)$ & $\mathcal{U}(-\pi,\,\pi)$ & $\mathcal{U}(-\pi,\,\pi)$\\[1mm]
Interpolation Power & { } & $\mathcal{U}(0.5,\,2)$ & $\mathcal{U}(0.5,\,2)$ & $\mathcal{U}(0.5,\,2)$ \\[1mm]
Blackbody ratio & { } & $\mathcal{N}(11,\,2)$ & $\mathcal{N}(260,\,15)$ & $\mathcal{N}(75,\,15)$\\[1mm]
Radius ratio & { } & $\mathcal{N}(12.1,\,1)$ & $\mathcal{N}(7.1,\,1)$ & $\mathcal{N}(11.75,\,1.5)$ \\[1mm]
{       } & {     } & $A_1$ = $\mathcal{N}(2.1,\,0.3)$ & $\mathcal{U}(-5,\,94)$ &  $\mathcal{U}(33,\,66)$\\[1mm]
{ }& { } & $A_2$ = $\mathcal{U}(-44.25,\,-27.25)$ & $\mathcal{U}(-48,\,-124)$ & $\mathcal{N}(6.8,\,1.7)$ \\[1mm]
{Secondary Effect Amplitudes}& ppm & $B_2$ = $\mathcal{U}(4.25,\,28.05)$ & $\mathcal{U}(-263,\,-363)$  &     \nodata \\[1mm]
{       }& {       } & $A_3$ = $\mathcal{U}(5.35,\,22.35)$ & \nodata  &  \nodata \\[1mm]
{       }& {       } & $B_3$ = $\mathcal{U}(-15.5,\,9.7)$ & \nodata &   \nodata \\[1mm]
\enddata
\tablecomments{The notation $\mathcal{N}(\mu,\,\sigma^{2})$ indicates a normal distribution with mean $\mu$ and standard deviation $\sigma$. The notation $\mathcal{U}(a,\,b)$ indicates a uniform distribution with a lower bound $a$ and upper bound $b$.}
\end{deluxetable*}

\begin{figure*}
\label{fig:model validation plots}
\gridline{\fig{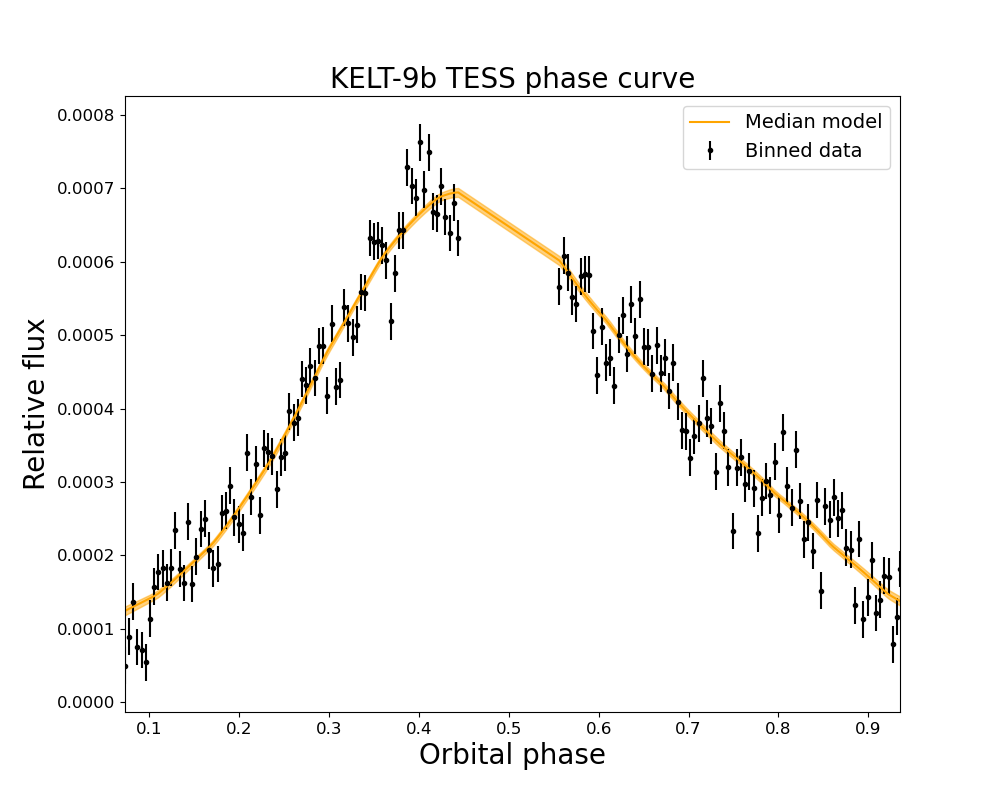}{0.5\textwidth}{\textbf{a)} TESS phase curve of KELT-9b modeled using our full phase curve model. Data was binned in 10 minute intervals shown in black.}
\fig{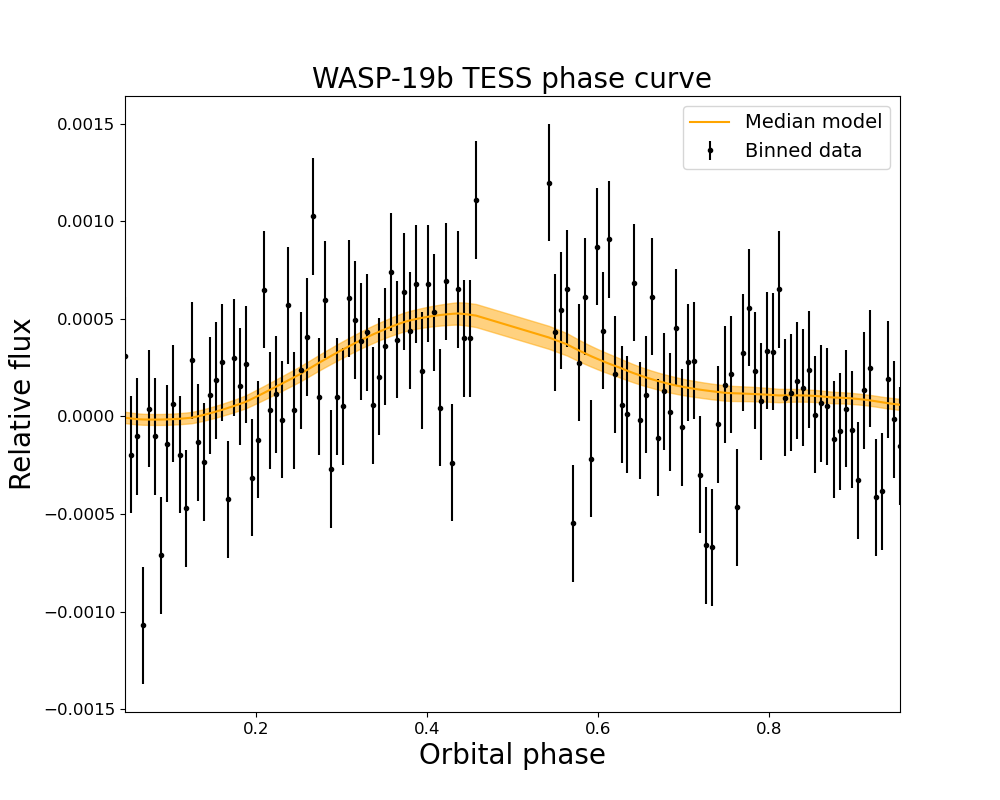}{0.5\textwidth}{\textbf{b)} TESS phase curve of WASP-19b modeled using our full phase curve model. Data was binned in 8 minute intervals shown in black.}}
\gridline{\fig{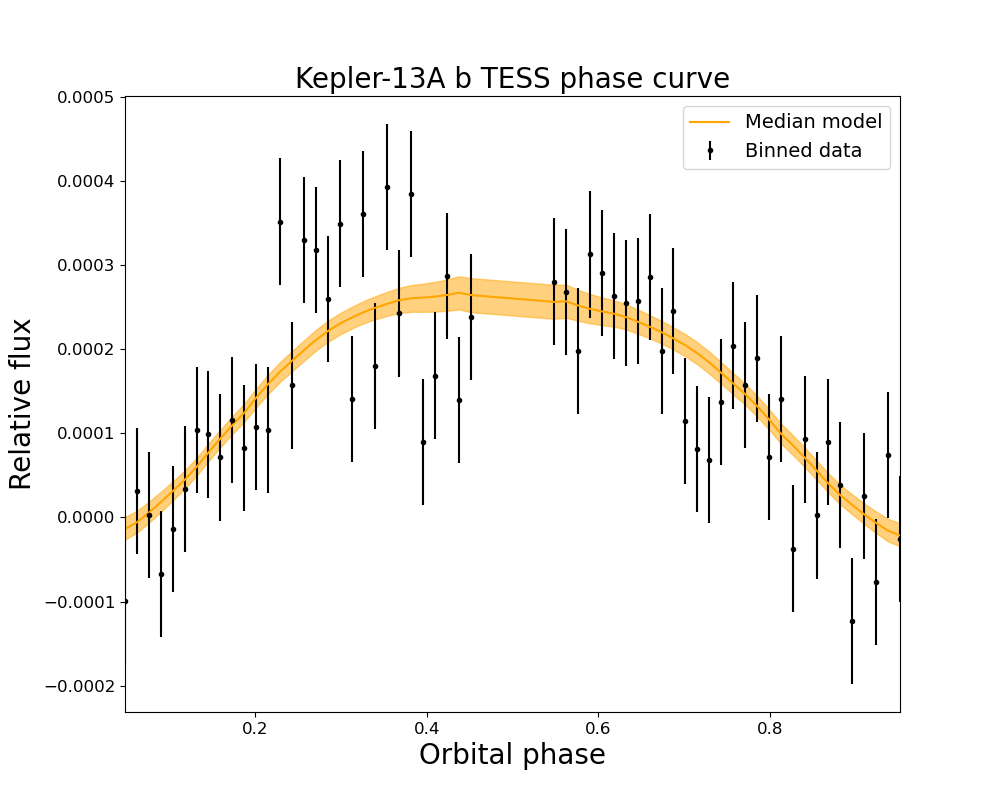}{0.5\textwidth}{\textbf{c)} TESS phase curve of Kepler-13Ab modeled using our full phase curve model. Data was binned in 35 minute intervals shown in black.}}
\caption{Phase curve models of TESS data for KELT-9b, WASP-19b and Kepler-13Ab. The median model was determined by calculating the 16th, 50th, and 84th percentile of the equal-weight posterior distribution of the models. This range is shown in orange.}
\end{figure*}

\subsection{KELT-9 b} \label{subsec:kelt9b test}

We fit our model to phase curve data for the ultra-hot Jupiter KELT-9b observed by the TESS telescope. Details of the data analysis and observations can be found in \citep{Wong2020kelt9}. Data was binned into 10 minute intervals following \citep{Wong2020kelt9}'s procedure. We used Equation 2 in \citep{Wong2020kelt9} to replace $f_{harmonics}$ in our full phase curve model (Eq. \ref{eq:fullphasecurve}).

\begin{figure*}
\hspace*{0.3cm} 
\includegraphics[scale=0.3]{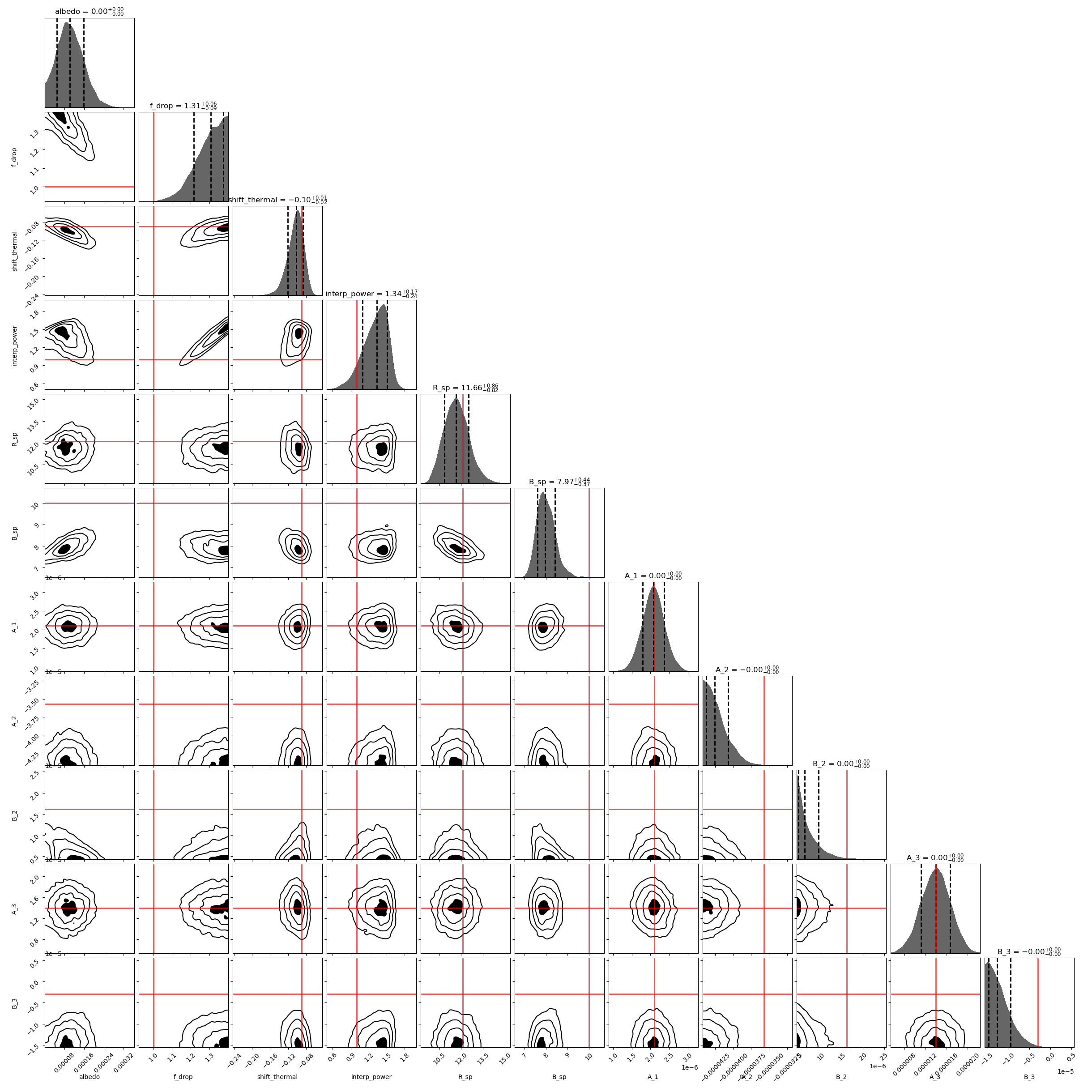}
\caption{Corner plot of posteriors for our model fit of KELT-9b TESS data. The red lines in the albedo, shift thermal, $A_1$, $A_2$, $B_2$, $A_3$, and $B_3$ correspond to the best-fit values from \citep{Wong2020kelt9}. The red lines in our other parameters correspond to the calculated, but not necessarily best-fit, values for the parameters in our model.
\label{fig:kelt9bcorner}}
\end{figure*}

We set uniform priors for all parameters except the radius ratio between the star and planet ($R_{sp}$), the blackbody radiation ratio between the star and planet ($B_{sp}$), and the Doppler boosting amplitude ($A_1$). A Gaussian prior was used for the $R_{sp}$ as it can be reasonably constrained by transit observations. The mean for the prior distribution was calculated from the stellar and planetary radii reported in \cite{gaudi2017}. The standard deviation for the prior distribution was based upon the maximum and minimum ratios calculated from the stellar and planetary radii reported in the NASA Exoplanet Archive. A Gaussian prior was also used for $B_{sp}$ as both $R_{sp}$ and $B_{sp}$ contribute to the planet-star flux ratio as shown in equation (Eq. \ref{eq:thermal2}) in our model. The mean for the prior distribution was based upon the calculated blackbody radiation ratio in the TESS bandpass. The standard deviation for the prior distribution was based upon the maximum and minimum ratios calculated from the stellar and planetary temperatures reported in the NASA Exoplanet Archive. A Gaussian prior was used for $A_1$ following \citep{Wong2020kelt9}'s procedure. For the other secondary effect amplitude parameters, we set the priors to vary within the error range of \citep{Wong2020kelt9}'s best-fit values. Prior bounds for albedo amplitude correspond to a geometric albedo between 0 and 1. The median model is shown in Figure \ref{fig:model validation plots}. Parameters converged reasonably well and posterior distributions for each parameter are shown in Figure \ref{fig:kelt9bcorner}. Best-fit values for albedo, amplitude and phase offset were within 1.8$\sigma$ and the secondary effect amplitudes were within 1.4$\sigma$ of \cite{Wong2020kelt9}'s results as shown in Table \ref{tab:modeltesting}.

\subsection{WASP-19 b} \label{subsec:wasp19b test}

In addition, we tested our model using phase curve data for the ultra-hot Jupiter WASP-19b observed by the TESS telescope. Details of the data analysis and observations can be found in \citep{Wong2020wasp19}. Data was binned into 8 minute intervals following \cite{Wong2020wasp19}'s procedure. For WASP-19b, we used Equation 2 in \cite{Wong2020wasp19} in place of $f_{harmonics}$ in our full phase curve model. 

Unlike KELT-9b, WASP-19b has a reported albedo of 0.16 $\pm$ 0.04 and is thus dominated by reflected starlight in the TESS bandpass \citep{Wong2020wasp19}. Initial modeling of the planet did not yield constrained results due to the thermal component parameters remaining unconstrained. We instead modeled the thermal and reflected components separately to determine if the thermal component of WASP-19b's emission could be constrained. We modeled the thermal component of the TESS data using only Equation \ref{eq:thermal} and the same priors as the joint fit for thermal parameters (flux drop $f_{drop}$, thermal offset ($\delta$), interpolation power ($i$), radius ratio ($R_{sp}$) and blackbody radiation ratio ($B_{sp}$)). The thermal component parameters remained unconstrained. This agrees with the previous analysis that the day-night temperature contrast, or thermal component, contribution is little and therefore can not be constrained by the data \citep{Wong2020wasp19}. All thermal component parameters reported in Table \ref{tab:modeltesting} are reported from our joint fit and contain large error bars due to the negligible amount of thermal contribution.

The same process was used to model the reflected light component using only Equation \ref{eq:reflection}. Albedo is the only reflected light component parameter as shown in Eq. \ref{eq:fullphasecurve} and remained constrained following our fit. This result further agrees with 
\cite{Wong2020wasp19}'s conclusion that the reflected light contribution dominates the phase curve. The albedo from our joint fit is reported in Table \ref{tab:modeltesting}. Our retrieved albedo is more elevated than the one reported in \cite{Wong2020wasp19}, but remains within 1.6 $\sigma$ of their results.

\begin{figure*}
\centering
\hspace*{0.3cm} 
\includegraphics[scale=0.35]{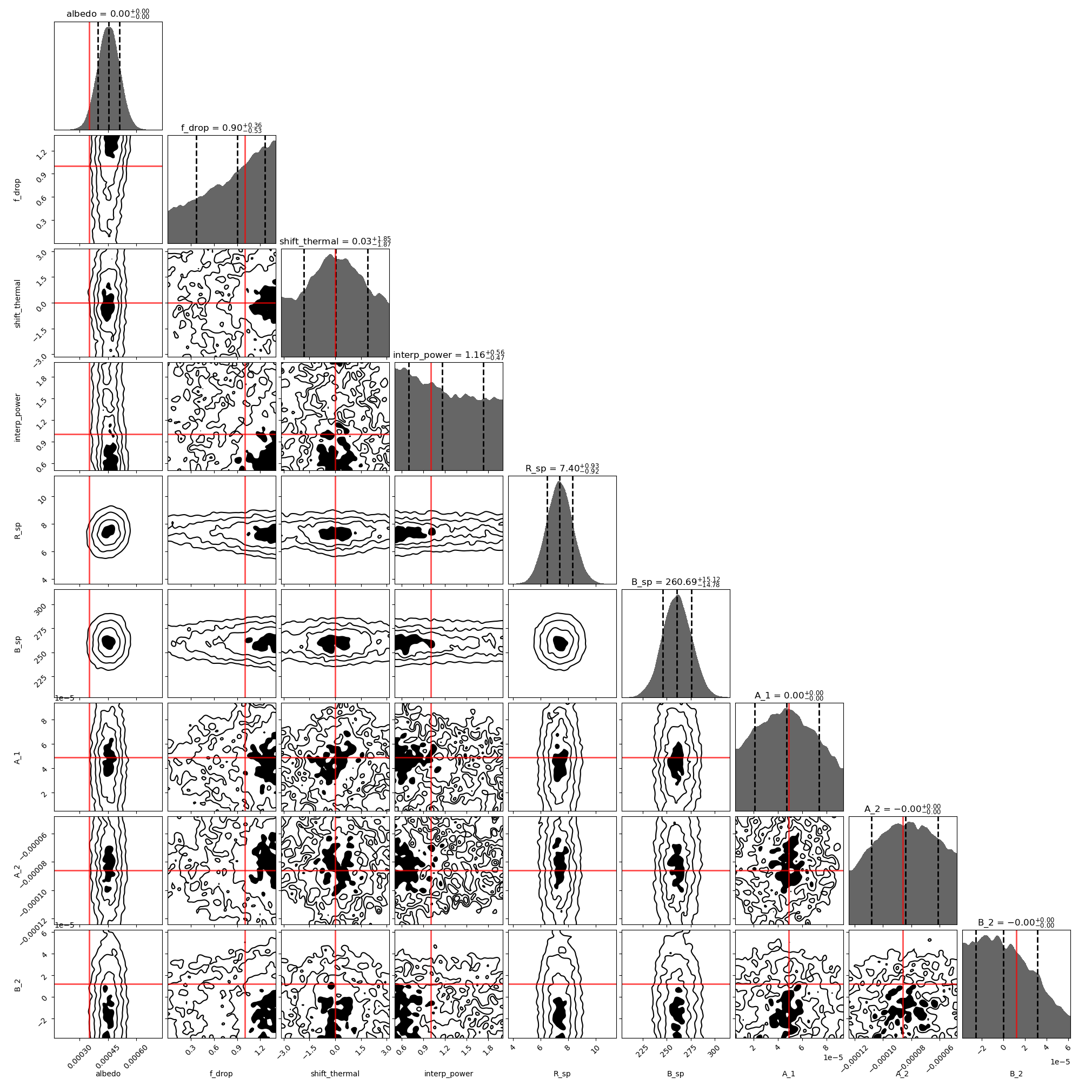}
\caption{Corner plot of posteriors for our model fit of WASP-19b TESS data. The red lines in the albedo, shift thermal, $A_1$, $A_2$, and $B_2$ correspond to the best-fit values from \citep{Wong2020wasp19}. The red lines in our other parameters correspond to the calculated, but not necessarily best-fit, values for the parameters in our model.
\label{fig:wasp19bcorner}}
\end{figure*}

All other parameters reported in Table \ref{tab:modeltesting} are from our joint fit. Just as with KELT-9b, we set uniform priors for all parameters except for $R_{sp}$ and $B_{sp}$. We followed the same procedure as KELT-9b when determining the priors for the albedo amplitude, blackbody radiation ratio and the secondary effect amplitudes. We show the posterior distributions of the joint fit in Figure \ref{fig:wasp19bcorner}. The reported values for albedo, amplitude and phase offset were within 1.6$\sigma$ and the secondary effect amplitudes were within 0.2$\sigma$ of \cite{Wong2020wasp19}'s reported error range.

\subsection{Kepler-13A b} \label{subsec:kepler13ab test}

Finally, we modeled TESS phase curve data from the hot Jupiter Kepler-13Ab, also known as KOI-13b. Details of the data analysis and observations can be found in \citep{Wong2021}. Data was binned into 35 minute intervals following \citep{Wong2021}'s procedure. For Kepler-13Ab, we used Equation 8 in \citep{Wong2021} in place of $f_{harmonics}$ in our full phase curve model.

Like WASP-19b, Kepler-13Ab has an elevated geometric albedo in the TESS bandpass: 0.53 $\pm$ 0.15, indicating that reflected light is the dominant component in the phase curve \citep{Wong2021}. Initial modeling of the planet did not yield constrained results due to the thermal component parameters remaining unconstrained. We instead modeled the thermal and reflected components separately following the same procedure as WASP-19b to determine if the thermal component of Kepler-13Ab's emission could be constrained.

The thermal component parameters remained unconstrained, indicating the thermal contribution is negligible compared to the reflected light contribution. This agrees with the elevated albedo reported in \cite{Wong2021}. All thermal component parameters reported in Table \ref{tab:modeltesting} are reported from our joint fit and contain large error bars due to this small contribution.

The same process was used to model the reflected light component using only Equation \ref{eq:reflection}. The albedo remained constrained following our fit. This result further agrees with our conclusion that the reflected light contribution dominates the phase curve. The albedo from our joint fit is reported in Table \ref{tab:modeltesting}. Our retrieved albedo is more elevated than the one reported in \cite{Wong2021}, but remains within 1 $\sigma$ of their results.

\begin{figure*}
\centering
\hspace*{0.3cm} 
\includegraphics[scale=0.37]{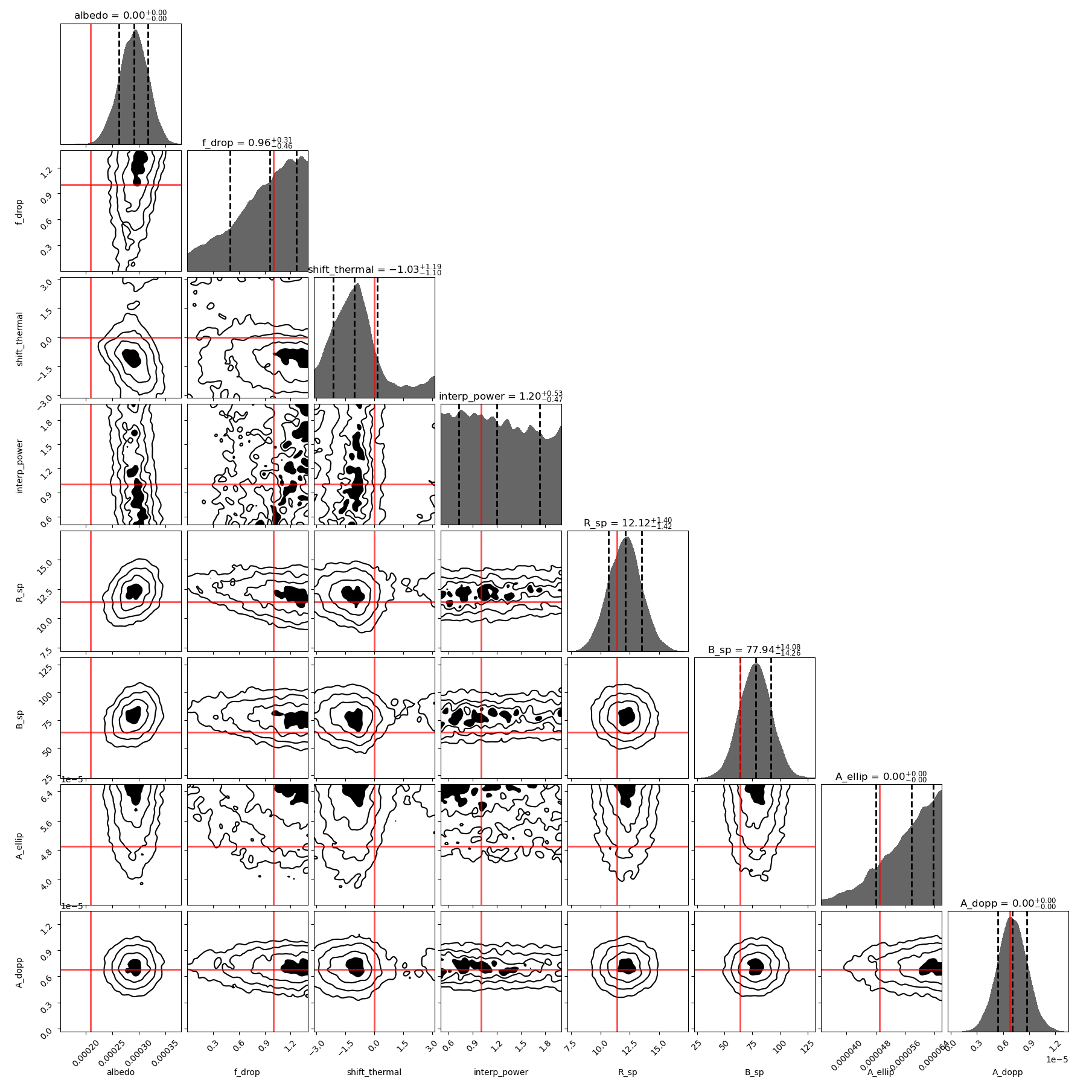}
\caption{Corner plot of posteriors for our model fit of Kepler-13Ab TESS data. The red lines in the albedo, shift thermal, $A_{dopp}$ and $A_{ellip}$ correspond to the best-fit values from \citep{Wong2021}. The red lines in our other parameters correspond to the calculated, but not necessarily best-fit, values for the parameters in our model.
\label{fig:kepler13acorner}}
\end{figure*}

All other parameters reported in Table \ref{tab:modeltesting} are from our joint fit. Just as with KELT-9b and WASP-19b, we set uniform priors for all parameters except for $R_{sp}$, $B_{sp}$ and $A_{dopp}$. We used a Gaussian prior for $A_{dopp}$ following \cite{Wong2021}'s procedure and used the same mean and standard deviation for the prior distribution. We followed the same procedure as our other analyses when determining the priors for the albedo amplitude, radius ratio, blackbody radiation ratio and the secondary effect amplitudes. We show the posterior distributions of the joint fit in Figure \ref{fig:kepler13acorner}. Best-fit values for albedo, amplitude and phase offset were within 1$\sigma$ and the secondary effect amplitudes were within 0.4$\sigma$ of \cite{Wong2021}'s reported error range.

\bibliography{sample631}
\bibliographystyle{aasjournal}
\end{CJK*}

\end{document}